\documentclass[a4paper,11pt]{article}
\pdfoutput=1 

\usepackage{jcappub} 

\usepackage[T1]{fontenc} 
\usepackage{textcomp}
\usepackage{gensymb}
\usepackage{siunitx}
\usepackage{physics}
\usepackage{aas_macros}
\usepackage{subcaption}
\usepackage{float}
\title{\boldmath Simultaneously unveiling the EBL and intrinsic spectral parameters of gamma-ray sources with Hamiltonian Monte Carlo}
\author[a,1]{M. Genaro,\note{Corresponding author.}}
\author[b]{L. A. Stuani Pereira,}
\author[c]{D. R. de Matos Pimentel,}
\author[a,1]{E. Moura Santos}

\affiliation[a]{Universidade de São Paulo - Instituto de Física, 05508-090, São Paulo (SP), Brazil}
\affiliation[b]{Universidade Federal de Campina Grande – Unidade Acadêmica de Física, 58429-900, Campina Grande (PB), Brazil}
\affiliation[c]{Parque Tecnológico São José dos Campos - Dom Rock, 12247-016, São José dos Campos (SP), Brazil}

\emailAdd{mgdantas@if.usp.br}
\emailAdd{luizstuani@uaf.ufcg.edu.br}
\emailAdd{douglas.roberto.fis@gmail.com}
\emailAdd{emoura@if.usp.br}

\abstract{The Extragalactic Background Light (EBL) is the main radiation field responsible for attenuating extragalactic gamma-ray emission at very high energies, but its precise spectral intensity is not fully determined. Therefore, disentangling propagation effects from the intrinsic spectral properties of gamma-ray sources (such as active galactic nuclei, AGN) is the primary challenge to interpreting observations of these objects. We present a Bayesian and Markov Chain Monte Carlo approach to simultaneously infer parameters characterizing the EBL and the intrinsic spectra in a combined fit of a set of sources, which has the advantage of easily incorporating the uncertainties of both sets of parameters into one another through marginalization of the posterior distribution. Taking a sample of synthetic blazars observed by the ideal CTA configuration, we study the effects on the EBL constraints of combining multiple observations and varying their exposure. We also apply the methodology to a set of 65 gamma-ray spectra of 36 different AGNs measured by current Imaging Atmospheric Cherenkov Telescopes, using Hamiltonian Monte Carlo as a solution to the difficult task of sampling in spaces with a high number of parameters. We find robust constraints in the mid-IR region while simultaneously obtaining intrinsic spectral parameters for all of these objects. In particular, we identify Markarian 501 (Mkn 501) flare data (HEGRA/1997) as essential for constraining the EBL above 30\,$\mu$m.}

\begin{document}
\maketitle
\flushbottom

\section{Introduction}

Since the mid-2000s, there has been a steady increase in the detection of extragalactic very-high-energy (VHE, $\gtrsim\SI{100}{GeV}$) sources \cite{2016Dermer}, mainly driven by the current generation of Imaging Atmospheric Cherenkov Telescopes (IACTs), represented by H.E.S.S. \cite{2006A&Ahess}, MAGIC \cite{2016APhmagic} and VERITAS \cite{2002APhveritas}. Currently, the TeVCat catalog \cite{2008tevcat} contains more than 70 extragalactic objects with measured gamma-ray spectra at those energies, most of them being blazars of the BL Lac type \cite{2022Galax..10...39B, 2021Mathieu}. With the next generation of IACTs, the Cherenkov Telescope Array (CTA) \cite{2013APhCTA}, many new VHE sources are expected to be discovered \cite{CTAConsortium:2017dvg}
due to CTA's improved sensitivity, extended energy coverage, and survey capabilities. To correctly interpret extragalactic VHE data, however, one must be aware of attenuation effects in the propagation of gamma rays with such energies. It is well established that gamma rays can interact with low energy background photons through pair production $\gamma\gamma \to e^+e^-$ (the Breit-Wheeler process \cite{Gould:1967zza, Gould:1967zzb}), which mostly happens with photons from the Extragalactic Background Light (EBL), the isotropic component of the extragalactic radiation (excluding the cosmic microwave background) in the wavelength range 0.1--1000\,$\si{\micro\meter}$ \cite{2013APhDwek}. The EBL is the direct result of star and structure formation histories, mostly constituted by starlight emission, some active galactic nuclei (AGN) and light reprocessed by dust, all redshifted by the cosmological expansion \cite{Hauser2001cosmic, 2016EBLmeasur}.

The exact EBL levels, on the other hand, are not precisely known, since direct measurements need a careful subtraction of foreground contamination (zodiacal light, atmospheric airglow, diffuse Galactic light, see e.g. \cite{Hauser2001cosmic, 2019ConPh..60...23M}). At the same time, galaxy counts can provide lower limits to the intensity \cite{2016EBLmeasur, 2021MNRAS.503.2033K}. Relevant tensions, however, have recently emerged from direct measurements with the New Horizons probe \cite{2022ApJ...927L...8L}, as they found an anomalous flux compared to the results from integrated light of external galaxies, further motivating alternative methods for probing the EBL.

Information about the intrinsic spectrum of extragalactic gamma-ray sources is able to constrain the EBL levels or inform upper limits. In general, this can be done by detailed modeling of the blazar's spectrum energy distribution (SED), by extrapolating high-energy measurements (not affected by attenuation) into the VHE domain or limiting the hardness of the spectra \cite{Costamante:2013sva}. The first method depends on the existence of simultaneous broadband SED observations (typically X-rays, but also radio and optical), which are not always available for every source detected in gamma rays. From this modeling, the EBL flux can be reconstructed \cite{Guy:2000xw}, or the optical depth associated with extragalactic attenuation can be derived when many sources at different redshifts are combined \cite{Dominguez:2013lfa}.  Other constraints on the spectra can produce upper limits on the EBL density, by excluding EBL parameters that generate unphysical spectra of blazars (e.g. \cite{HESS:2005ohe, Dwek2005ApJ, Meyer2012A&A}). These limits may require a lower bound to the spectral index, such as $\Gamma \gtrsim 1.5$, if the photons are produced by inverse Compton scattering \cite{Mazin:2007pn}, or assume the absence of pile-up and exponential rise features at the end of the spectrum \cite{Costamante:2013sva, Dwek2013APh}.
Such constraints can be achieved in a ``model-independent'' way, by exploring a grid of generic EBL shapes \cite{Meyer2012A&A, Mazin:2007pn} and decoupling the redshift evolution from the local EBL density \cite{2015ApJBiteau}, or using established EBL models (e.g. \cite{2010finke, dominguez2011extragalactic, 2012MNRAS.422.3189G}), and scaling their optical depths to find best-fit conditions \cite{2013A&A...550A...4H, Desai:2019pfl}. An iterative procedure can also be used to choose the intrinsic spectral model, as in \cite{2019A&A...627A.110B}, avoiding possible bias and increase the robustness of the constraint.

All approaches naturally benefit from a larger sample of data, particularly sources at distinct redshifts to probe EBL levels and its evolution. A simultaneous fit of the data is also key for breaking degeneracies in the EBL description, as even though each source may have its own intrinsic spectrum, a single EBL model should provide the appropriate attenuation to all sources at different redshifts. In \cite{2018Scifermi}, the Fermi-LAT collaboration was able to estimate the optical depth due to the EBL using a sample of 739 active galaxies between $0.03 < z < 3.1$, obtaining an EBL spectrum at $z=0$ close to the measurements from galaxy counts. However, the energy coverage of Fermi implies constraints only up to the near-IR at $\SI{4}{\micro\meter}$ (or even smaller wavelengths for higher redshift). In fact, data from IACTs, reaching a few tens of TeV, are essential for constraining the mid-IR ($3$--$\SI{25}{\micro\meter}$), as the cross-section of pair-production peaks at an EBL wavelength $\lambda (\si{\micro\meter})\sim 1.24E_\gamma (\si{\tera\electronvolt})$ for a gamma-ray photon detected at energy $E_\gamma$ \cite{Hauser2001cosmic}. Yet, current IACT data, as identified by \cite{2019pimentel} using a sample cataloged in TeVCat, are heavily dominated by spectra whose attenuation originates from the stellar component of the EBL (mainly UV, optical light and near-IR). Consequently, further observations reaching higher energies or performed with improved precision, such as expected with the CTA, will refine the determination of the infrared EBL spectrum and the mid-IR portion in particular, where direct measurements are heavily affected by foreground contamination \cite{2005Kashlinksky}. In \cite{CTA:2020hii}, the CTA collaboration has performed forecast studies on the capabilities of the observatory to constrain the overall normalization of the EBL spectrum using synthetic gamma-ray spectra representing known TeV emitters, for which the redshifts are known a priori. The results show that in the redshift range $z<2$, the EBL normalization can be determined with a statistical uncertainty below 15\%. The correct distinction of intrinsic and EBL attenuation features is also key for producing constraints on new physics, such as signs of Lorentz invariance violation \cite{Lang:2018yog, sym12081232} or coupling to axion-like particles, both topics to be greatly explored by CTA \cite{CTA:2020hii}.

This work aims at contributing to the task of disentangling intrinsic spectrum and absorption features from data, by applying Monte Carlo methods and Bayesian inference to a sample of gamma-ray spectra from TeV emitters. The Bayesian approach should provide us with the joint probability distribution of parameters characterizing the EBL and intrinsic spectra of gamma-ray sources conditioned to the observed data, from which any desired statistics and credible intervals can be computed. 

We present our study in two parts. Firstly, we simulate observations of blazars under CTA's ideal telescope configuration (described in section \ref{sec:source_selection}) and select sources with high detection significance to constrain intrinsic and EBL parameters. In this controlled scenario, we study the resolution of EBL parameters as a function of the number of sources and observation campaign duration. The synthetic samples also allow us to explore some systematic effects associated with the modeling of the EBL density. Secondly, we apply our methodology to a set of 65 spectra from 36 extragalactic gamma-ray sources measured by current IACTs, investigating the resulting EBL constraints while simultaneously determining all their intrinsic parameters. For this analysis, the Hamiltonian Monte Carlo approach was very important to reach an efficient sampling in a parameter space with high dimensionality. Section \ref{sec:methodology} gives detailed information on the methodology, including descriptions of the EBL model, the spectral data, and optimisations to the Monte Carlo sampler, respectively in sections \ref{sec:ebl_model}, \ref{sec:source_selection} and \ref{sec:marginal}. This is followed by the analysis of EBL and intrinsic parameters constraints using simulated data in section \ref{sec:results_simulated} and using observed data in section \ref{sec:tevcat}. Finally, conclusions are drawn in section \ref{sec:conclusion}.
 
\section{Methodology}\label{sec:methodology}

Using Monte Carlo methods and Bayes' theorem, we can sample the posterior probability density distribution $p(\boldsymbol{\Omega}|D,I)$ of parameters $\boldsymbol{\Omega}=\left\{\omega_{\textrm{EBL}},\omega_\text{S}\right\}$ modeling the EBL ($\omega_\text{EBL}$) and the gamma-ray spectra ($\omega_\text{S}$), given the available data $D$ and possibly extra information $I$,

\begin{equation}
p(\vb{\Omega}|D,I) = \frac{p(D|\vb{\Omega},I)p(\vb{\Omega}|I)}{p(D|I)}.
\end{equation}
The information $I$ represents, for instance, the choice of models to describe the intrinsic spectra of sources or the EBL attenuation. Any a priori information on $\vb{\Omega}$ may be incorporated in the prior probability distribution $p(\vb{\Omega}|I)$, but here we will only work with an uninformative prior $p(\boldsymbol{\Omega})=\textrm{constant}$, described in section \ref{sec:ebl_model}. When sampling the posterior with MCMC, we are reconstructing the joint probability distribution --- up to a normalization factor ---, which can be later marginalized over parts of the parameter space, allowing the computation of expectation values or credible intervals for selected parameters. In particular, by considering $\omega_S$ as nuisance parameters and integrating over them, the marginal distribution of EBL parameters can incorporate the statistical uncertainties of the unknown intrinsic spectra of the sources. Conversely, intrinsic source parameters can be estimated while marginalizing over possible EBL configurations.

Under the hypothesis of Gaussian errors, the observed flux is assumed to follow a normal distribution. Considering independent observations of $N$ gamma-ray sources, each containing  $n_j$ measured flux points with uncertainty $\sigma$, the likelihood can be expressed as
\begin{equation}\label{eq:likelihood}
    p(D|\vb{\Omega},I) = \frac{1}{Z}\exp{-\frac{1}{2}\sum^N_{j=1}\sum^{n_j}_{i=1}\left[\frac{\phi^{(j)}_{\rm obs}\left(E^{(j)}_i\right)-\phi^{(j)}_{\rm mod}\left(E^{(j)}_i;\vb{\Omega}\right)}{\sigma\left(E^{(j)}_i\right)}\right]^2},
\end{equation}
where $Z$ is the probability normalization factor and $E^{(j)}_i$ are the observed energy bins. The observed differential flux $\phi_\text{obs}$ is presented in units of $\si{\tera\electronvolt^{-1}\meter^{-2}\second^{-1}}$, while the modeled flux, $\phi_{\rm mod}$, can then be expressed in terms of the intrinsic\footnote{Here, we use the intrinsic nomenclature to refer to the emitted flux at the source, but no redshift correction is applied to the energy.} spectrum attenuated by the EBL opacity as
\begin{equation}
    \phi_{\rm mod}(E;\vb{\Omega}) = e^{-\tau(E;\omega_{\rm EBL})}\phi_{\rm intr}(E;\omega_\text{S}).
\end{equation}
The extragalactic attenuation is quantified by the optical depth $\tau$, which is energy and redshift dependent, expressed in terms of the EBL photon number density $n$ by 
\begin{equation}\label{eq:tau}
    \tau(E_\gamma,z)=c\int\limits^z_0\dd{z'}\abs{\dv{t}{z'}}\int\limits^1_{-1}\dd{\mu}\frac{(1-\mu)}{2}\int\limits^\infty_{E'_{\rm min}}\dd{E'}\sigma_{\text{BW}}(E',E'_\gamma,\mu)\,n(E',z';\omega_{\rm EBL}),
\end{equation}
where $\sigma_{\textrm{BW}}$ is the Breit-Wheeler cross-section and $\mu=\cos\theta$ is the collision angle between the gamma-ray and EBL photons in the proper reference frame (i.e., the reference frame of the interaction), with energies $E'_\gamma$ and $E'$ (primed quantities), respectively. Then, observed (unprimed) quantities, such as the gamma-ray energy, are redshifted by cosmological expansion, e.g. $E'_\gamma =(1+z)E_\gamma$. The integration limits consider all energies and angles allowed by the kinematics, where $E'_\text{min} = 2m^2c^4/[E'_\gamma(1-\mu)]$, with $m$ being the electron mass. Finally, the cosmological distance element is given by
\begin{equation}
    c\abs{\dv{t}{z}}=\frac{1}{1+z}\frac{c}{H_0}\frac{1}{\sqrt{\Omega_m(1+z)^3+\Omega_\Lambda}},
\end{equation}
assuming, for the current analysis, a $\rm \Lambda CDM$ cosmology in a spatially flat universe with $\Omega_m=0.3$, $\Omega_\Lambda = 0.7$ and $H_0=70\,\rm km\,s^{-1}\,Mpc^{-1}$, where the radiation density is negligible considering the redshifts of interest ($z < 1$). 

\subsection{EBL model}\label{sec:ebl_model}

The adopted EBL description, elaborated by \cite{2010finke} (hereinafter F10), is built through integration over star formation rates and stellar evolution properties, which creates the primary source of emission in the optical wavelengths, and then is partially absorbed by dust and re-emitted in the IR. Here, we focus on investigating constraints on the IR range of the EBL by changing the proportions of the dust constituents. In this model, three dust components are considered, each one emitting as a blackbody with fixed temperature: warm large dust grains (LG) at $\SI{40}{\kelvin}$, hot small grains (SG) at $\SI{70}{\kelvin}$ and a $\SI{450}{\kelvin}$ component representing polycyclic aromatic hydrocarbons (PAH). Therefore, the emissivity from dust is a weighted sum of the three blackbody components, normalized by the fraction of stellar comoving luminosity $j^{\rm star}$ absorbed in the interstellar medium.  That is,
\begin{equation}
    \epsilon j^{\rm dust}(\epsilon,z) = \frac{15}{\pi^4}\int\dd{\tilde{\epsilon}}\left[ \frac{1}{f_{\rm esc}(\tilde{\epsilon})}-1\right]j^{\rm star}(\tilde{\epsilon},z)\times \sum^3_{n=1}\frac{f_n}{\Theta^4_n}\frac{\epsilon^4}{\exp(\epsilon/\Theta_n)-1},
\end{equation}
where the fraction $f_{\rm esc}(\epsilon)$ of starlight photons with given energy $\epsilon = E/(m_ec^2)$ that escapes the galaxies was parameterized according to \cite{2009Razzaque}. The quantities $f_n$ are fractions of the absorbed emissivity re-emitted in each dust component, while $\Theta_n= k_BT/(m_ec^2)$ are their temperatures in units of the electron rest energy. Collectively, the dust fractions constitute the EBL parameters $\omega_{\rm EBL}$ of our interest.

The total emissivity $j(\epsilon,z)= j^{\rm star}(\epsilon,z)+j^{\rm dust}(\epsilon,z)$ must be integrated over the cosmological evolution to obtain the comoving energy density of the EBL. Since the dust fractions $f_n$ can be factored out of the emissivity, the optical depth can be written as a linear combination of the attenuation due to the stellar and three dust components
\begin{equation}
    \tau(E_\gamma,z) = \tau^{\rm star}+f_{\rm PAH}\tau^{\rm dust}(\Theta_{\rm PAH}) + f_{\rm SG}\tau^{\rm dust}(\Theta_{\rm SG})+f_{\rm LG}\tau^{\rm dust}(\Theta_{\rm LG}),
\end{equation}
showing explicitly the temperature dependence and omitting the $E_\gamma,z$ variables for notation simplicity. Furthermore, because the dust emissivity is computed self-consistently with the fraction of absorbed stellar radiation, the dust fractions are normalized imposing the condition
\begin{equation}
    f_{\rm PAH}+f_{\rm SG}+f_{\rm LG} = 1.
\end{equation}
This is incorporated in our analysis by choosing to set $f_{\rm LG}=1-f_{\rm PAH}-f_{\rm SG}$ during the MCMC sampling and is accompanied by an uniform prior which imposes $0\le f_{\rm PAH}+f_{\rm SG} \leq 1$.

When working with the simulated sources, we use the EBL grid (i.e., the optical depth in the $E_\gamma\times z$ parameter space) as computed by \cite{2019pimentel}. 
However, for analyzing real IACT flux data, we have recomputed the optical depth using a larger range in redshift. The description of both grids can be found in table~\ref{tab:grids_info}, while table \ref{tab:true_values_ebl} presents the reference values of the EBL parameters as used throughout the analysis.

\begin{table}[t]
    \centering
    \begin{tabular}{c|c|c|c|c}
    \hline
        Grid & Variable & Range & No. of points & Spacing \\\hline\hline
        \cite{2019pimentel} & Energy &$0.01$--$\SI{100}{\tera\electronvolt}$ &100 & log \\
         & Redshift & $0.01$--$6.00$ & 600 & linear\\\hline
         \cite{2019pimentel} & Energy &$0.01$--$\SI{100}{\tera\electronvolt}$ & 50 & log \\
         (Updated) & Redshift & $10^{-4}$--$1$ & 50 & log\\\hline
    \end{tabular}
    \caption{Description of the EBL optical depth grids, based on the F10 model. We use the grid computed by \cite{2019pimentel} to analyse the synthetic sources (section \ref{sec:results_simulated}), while having an updated grid used for the IACT data inference (section \ref{sec:tevcat}).}
    \label{tab:grids_info}
\end{table}

\begin{table}[t]
    \centering
    \begin{tabular}{c||c|c|c||c|c|c}
    \hline
        Variable & $f_{\text{PAH}}$ & $f_{\text{SG}}$ & $f_{\text{LG}}$ & $T_\text{PAH}$ & $T_\text{SG}$ & $T_\text{LG}$\\\hline
        \cite{2019pimentel} value & 0.25 & 0.05 & 0.70 & $\SI{450}{\kelvin}$ & $\SI{70}{\kelvin}$ & $\SI{40}{\kelvin}$\\\hline 
    \end{tabular}
    \caption{Reference values of the parameters describing the EBL density of the F10 EBL model implemented by \cite{2019pimentel}  (dust fractions and their respective temperatures).}
    \label{tab:true_values_ebl}
\end{table}

\subsection{Selection of real and synthetic sources and spectral models}\label{sec:source_selection}

For our studies using synthetic data, a source population of BL Lacs was sampled according to a luminosity function tuned in the GeV energy range to the Fermi-LAT data (1FGL) \cite{Ajello:2013lka}. The AGN spectra were then extrapolated to the TeV region (from $\SI{100}{\giga\electronvolt}$ to $\SI{100}{\tera\electronvolt}$) assuming a power-law spectral shape at the source. We also accounted for the absorption of the VHE gamma-ray flux due to the interaction with the EBL by using F10 and EBL emissivity model from \cite{dominguez2011extragalactic} (hereinafter D11). The observations of the synthetic BL Lacs were simulated to be consistent with those of the future CTA observatory using the CTOOLS software framework \cite{ctools}, taking into account the instrument properties  for the full-scope configuration of the northern and southern arrays (CTA instrument response function \texttt{prod3b-v2} \cite{cherenkov_telescope_array_observatory_2016_5163273}). In this configuration, the northern array has 4 Large-Sized Telescopes (LSTs) and 15 Medium-Sized Telescopes (MSTs), while the southern array contains 4 LSTs, 25 MSTs and 70 Small-Sized Telescopes (SSTs). The current up-to-date instrument response function is the \texttt{prod5-v0.1} \cite{cherenkov_telescope_array_observatory_2021_5499840}, which corresponds to the expected performance of CTA during its first construction phase. In this ``Alpha'' configuration, the northern array has 4 LSTs and 9 MSTs, while the southern array has no LSTs, 14 MSTs and 37 SSTs, besides a slightly different disposition of the telescopes. The specific differences in layout (including sensitivity curves) can be seen in the downloadable material available in references \cite{cherenkov_telescope_array_observatory_2016_5163273} and \cite{cherenkov_telescope_array_observatory_2021_5499840}. Naturally, there is a loss of sensitivity by having fewer telescopes, but in both cases the flux sensitivity is best at the 0.5--$\SI{20}{\tera\electronvolt}$ energy range, which should result from the presence of MSTs in these configurations. This range is also the most relevant for this work, given the focus on the interaction with the EBL. Therefore, as more telescopes are incorporated into the actual arrays, the improved sensitivity at VHEs should provide refined data for EBL constraints, although similar effects can be accomplished by increasing the observation time when available. Here we present one optimistic scenario with the full-scope configuration, but further studies are required to quantify the impacts from other array dispositions.

The survey on the population of BL Lacs is performed using the normal and conservative pointing mode, in which each telescope is pointed to the same location in the sky covering an area of about $50.6\,\text{deg}^2$ at once, and the observations were optimized for low zenith angles (smaller than 45\degree). A region of about 25\% of the sky, 5\degree\, above the galactic equator and $-90\degree<l_\text{gal}<90\degree$, was covered with pointings uniformly spaced with the help of HEALPix \cite{healpix} (3.16$^\circ$ spacing between adjacent pointings with $\texttt{nside}=16$). This strategy should provide a reasonably realistic distribution of offsets between the source positions and the telescope axis, an important parameter affecting the significance of the detection. Given the assumptions on the luminosity function and spectral models of the synthetic sample, we do not aim at making a forecast for the future operation of CTA from these simulated observations, but simply characterize the performance of the EBL reconstruction method. Readers may be interested in checking the studies on the extragalactic TeV source population expected with CTA \cite{2017ICRC...35..632H}, which are based on the 3FHL catalog \cite{2017ApJS..232...18A} and provide distinct flux extrapolations of the source's emission.

All sources had their detection significance computed considering the test statistic defined by
\begin{equation}
    TS = -2(\ln{\mathcal{L}_0} - \ln{\mathcal{L}_1}),
\end{equation}
where $\mathcal{L}_0$ is the likelihood of photon detection under the null hypothesis of pure background (i.e. cosmic rays), while $\mathcal{L}_1$ refers to the alternative hypothesis in which a source is also present. 

For sources with $TS > 25$, we have simulated 5\,h of extra observation time per pointing to get improved statistics representing longer observational campaigns. Therefore, we have separated the data into two groups of statistically significant source detections: one consisting of spectra measured with up to 5 hours of total observation time and another of spectra measured 5\,h or more (both sets with multiple pointings per source). We refer to these groups as the ``short'' and ``long'' observational time, respectively. 

Finally, we analyse a sample of real sources catalogued in TeVCat (whose publicly available SEDs have been previously collected by \cite{2019pimentel}). From this selection, we eliminated sources with highly uncertain (or unknown) redshift, since this could introduce a potential bias in the optical depth. Also, to guarantee a minimum fit quality, we have only incorporated spectra with four or more flux points. This resulted in a sample of 65 spectra from 36 distinct sources, mainly comprised of BL Lacs of the high-frequency peak type (HBL) with $z < 1$. The complete list of sources is presented in table~\ref{tab:tevcat_sources} and section~\ref{sec:tevcat} discusses their analysis. The local sources ($z < 0.02$) are not expected to provide relevant EBL constraints, but their inclusion is not biasing our analysis according to the tests we performed, so they are kept in the sample.

\begin{table}[tbp]
    \centering
    \begin{tabular}{c|c|c||c|c|c}
    \hline
    \hline
        Name & $z$ & Type & Name & $z$ & Type \\
        \hline
        1ES 0229+200 & 0.14 & HBL & Centaurus A & 0.00183 & FR-I \\
        1ES 0347--121 & 0.188 & HBL & H 1426+428 & 0.129 & HBL \\
        1ES 0414+009 & 0.287 & HBL & H 2356--309 & 0.165 & HBL\\
        1ES 0806+524 & 0.138 & HBL & IC 310 & 0.0189 & AGN (unknown)\\
        1ES 1011+496 & 0.212 & HBL & M87 & 0.0044 & FR-I\\
        1ES 1101--232 & 0.186 & HBL & Markarian 180 & 0.045 & HBL\\
        1ES 1215+303 & 0.13 & HBL & Markarian 421 & 0.031 & HBL\\
        1ES 1218+304 & 0.182 & HBL & Markarian 501 & 0.034 & HBL\\
        1ES 1312--423 & 0.105 & HBL & NGC 1275 & 0.017559 & FR-I\\
        1ES 1727+502 & 0.055 & HBL & PKS 0447--439 & 0.343 & HBL\\
        1ES 1959+650 & 0.048 & HBL & PKS 1441+25 & 0.939 & FSRQ\\
        1ES 2344+514 & 0.044 & HBL & PKS 1510--089 & 0.361 & FSRQ\\
        1RXS J101015.9 & 0.142639 & HBL & PKS 2005--489 & 0.071 & HBL\\
        3C 279 & 0.5362 & FSRQ & PKS 2155--304 & 0.116 & HBL\\
        3C 66A & 0.34 & IBL & RBS 0413 & 0.19 & HBL\\
        4C +21.35 & 0.432 & FSRQ & RGB J0152+017 & 0.08 & HBL\\
        AP Librae & 0.049 & LBL & RGB J0710+591 & 0.125 & HBL\\
        BL Lacertae & 0.069 & IBL & RX J0648.7+1516 & 0.179 & HBL \\
        \hline\hline
    \end{tabular}
    \caption{Distinct gamma-ray sources selected to constrain EBL and intrinsic spectral parameters. Both redshift and object type were extracted from TeVCat \cite{2008tevcat}.}
    \label{tab:tevcat_sources}
\end{table}

Throughout this work, we adopt up to three different descriptions of the sources' intrinsic differential flux: a power law (PL), a log parabola (LP), and a power law with exponential cutoff (PLC). That is,
\begin{equation}\label{eq:intrinsic}
    \phi_{\rm intr}(E) = \begin{cases}
    N_0\left(\frac{E}{E_0}\right)^{-\Gamma} &\mbox{(PL)}\\
    N_0\left(\frac{E}{E_0}\right)^{-a-b\log(E/E_0)} &\mbox{(LP)}\\
    N_0\left(\frac{E}{E_0}\right)^{-\Gamma}e^{-E/E_{\rm cut}} &\mbox{(PLC)},
    \end{cases}
\end{equation}
where $E_0=\SI{1}{\tera\electronvolt}$ is fixed. Our simulated sample was exclusively generated with the PL model, but for the sources in TeVCat, we have also adopted LP and PLC ones. The process of choosing the intrinsic spectral model for each source is described in section~\ref{sec:tevcat}.

\subsection{MCMC sampler for synthetic sources}\label{sec:marginal}

The goal of the Bayesian inference is to reconstruct the posterior distribution of $\omega_{\rm EBL}=\{f_{\rm PAH}, f_{\rm SG}, f_{\rm LG}\}$, alongside intrinsic parameters (two or three for each source, depending on the adopted model). To analyse the synthetic sources, we have used the MCMC ensemble sampler \texttt{emcee} \cite{2013emcee}, a \texttt{python} affine-invariant algorithm\footnote{Available in \url{https://emcee.readthedocs.io/en/stable/}.} based on \cite{2010goodman}. The program creates an ensemble of parallel chains (called walkers), evolving them in a way analogous to the Metropolis-Hastings algorithm, by sampling a proposal step and using an acceptance rule to advance the Markov chain. However, the chains in the \texttt{emcee} method are not independent, as the algorithm divides the ensemble into two subsets and uses the positions of one set to evolve the other. This has the advantage of allowing parallel computing, speeding up the process. The standard ``move'' (how the new steps of the chain are proposed) is called the stretch move \cite{2010goodman}, but the current version of the code allows different proposal algorithms. From our test, the kernel-density-based proposal resulted in lower autocorrelation values, so it became the preferred configuration.

Naturally, the dimensionality of the parameter space grows linearly with the number of gamma-ray sources, adding computational time to the MCMC simulation. We succeeded in reducing this dimensionality by analytically marginalising over the flux normalisation variable, $N_0$, of each source. To do this, one needs to assume independence between the flux level and other parameters (e.g. spectral index). Although it is known that some blazars present a spectral hardening with increased flux levels, this study will not incorporate variability, therefore, for a given source observational campaign, we shall treat its spectral index as constant. The resulting marginalised likelihood, with one or two free intrinsic parameters for each source spectrum, can be found in appendix \ref{appendix:marginal}, alongside its derivation. In Appendix \ref{appendix:ebl_params} we also show the statistical properties of the simulations performed.

\subsection{Hamiltonian Monte Carlo} \label{sec:hmc}

To analyse the large amount of TeVCat sources we have opted to apply a branch of MCMC methods called Hybrid or Hamiltonian Monte Carlo (HMC) \cite{2011hmcm.book..113N}. This approach uses information about the geometry of the desired target distribution to more efficiently explore its typical set, which is the most important region to sample when computing expectation values. This is done by combining a stochastic exploration of the parameter space (as usually performed by MCMC methods) with deterministic trajectories obtained through solving Hamilton's equations. Thus, the original problem of sampling the target distribution $p(\vb{\Omega}|D,I)$ is mapped to an equivalent mechanical system, in which, along with our generalized coordinates $q=\vb{\Omega}$, we introduce conjugate momenta $P$ and define
\begin{equation}
 U = -\ln(p(q|D,I)),   
\end{equation}
as the corresponding potential energy. To perform the sampling, at each step of the HMC implementation the momentum values are randomly chosen, determining the ``energy'' level in which the conservative Hamilton dynamic takes place. After integrating Hamilton's equations for a number of specified steps, the new end coordinates become the proposal step evaluated by a modified Metropolis-Hastings acceptance criteria (see  \cite{2017hmc_conceptual} for more details). Such procedure scales very well at high dimensions, but requires a crucial tuning of some parameters to achieve efficient sampling (the choice of kinetic energy and ``masses'', besides the number and size of steps during Hamilton's dynamics).   

In our code, we utilize an Euclidean-Gaussian kinetic energy 
\begin{equation}
  K=\frac{1}{2}P^TM^{-1}P, 
\end{equation}
with a diagonal mass matrix $M$ estimated from the posterior distributions of test runs. To solve Hamilton's equations, we used a fourth-order Forest-Ruth symplectic integrator \cite{1990PhyD...43..105F} and have found that $L = 80$ integration steps with size $\Delta t = \num{8e-3}$ resulted in an acceptance fraction around $0.75$. In Appendix \ref{appendix:ebl_params}, we present the effective sample size and precision error of the final HMC simulations.

\section{Assessing accuracy and precision with synthetic samples}\label{sec:results_simulated}

We begin by investigating the ideal scenario, in which no systematic uncertainties are present in the modeling of EBL absorption or intrinsic spectra of the sources. To quantify the impact of a combined fit of blazar spectra, we have ordered the data according to their TS values and progressively included them in the analysis --- starting with the two highest TS sources. Considering blazars described by the PL model and attenuated by F10 EBL model, as implemented by \cite{2019pimentel} (grid and parameter values in tables \ref{tab:grids_info} and \ref{tab:true_values_ebl}), we perform two sets of MCMC simulations. These simulations take separately the subsets of short and long total observational time, with the goal of understanding how the increased observation period (and consequently more well-measured spectra) impacts the accuracy and precision of inferred parameters. In table~\ref{tab:sources} the characteristics of each source (redshift and spectral index) and of the simulated observation (TS and total time) are described.

\begin{table}[tbp]
    \centering
    \begin{tabular}{|c|c|c|c|c||c|c|c|}
    \hline
         \rule{0pt}{2.3ex}$z$ & $\Gamma_{\text{true}}$ & Obs. T. (h) & $E_\text{max}$(TeV) & TS & Obs. T. (h) &  $E_\text{max}$(TeV) & TS \\
         \hline
        \rule{0pt}{3ex}0.051 & 1.578 & 0.98 & $21.13$ & 15528  & 5  & $29.85$ & 85362\\[5pt]
        \rule{0pt}{2ex}0.038 & 1.873 & 1.96 & $21.13$ & 15499 & 5 & $21.13$ & 39116 \\[5pt]
        \rule{0pt}{2ex}0.058 & 1.587 & 3.92 & $14.96$ & 12095 & 20 & $21.13$ & 48921 \\[5pt]
        \rule{0pt}{2ex}0.061 & 1.714 & 3.92 & $14.96$ & 10250 & 20 & $14.96$ & 44412 \\[5pt]
        \rule{0pt}{2ex}0.079 & 1.485 & 0.98 & $14.96$ & 9691 & 5 & $21.13$ & 51800 \\[5pt]
        \rule{0pt}{2ex}0.076 & 1.477 & 0.98 & $12.59$ & 4081 & 20 & $21.13$ & 19381 \\[5pt]
        \rule{0pt}{2ex}0.083 & 1.899 & 2.21 & $12.59$ & 2887  & 15 & $10.59$ & 9236\\[5pt]
        \rule{0pt}{2ex}0.125 & 1.503 & 1.96 & $5.31$ & 2054 & 10 & $5.31$ & 11936 \\[5pt]
        \rule{0pt}{2ex}0.192 & 1.705 & 3.92 & $2.66$ & 1949 & 20 & $10.59$ & 9483 \\[5pt]
        \rule{0pt}{2ex}0.077 & 1.623 & 3.92 & $14.96$ & 1701 & 20 & $14.96$ & 7070 \\[5pt]
        \rule{0pt}{2ex}0.092 & 1.514 & 3.92 & $10.59$ & 1316 & 20 & $10.59$ & 5811 \\ [5pt]
        \rule{0pt}{2ex}0.076 & 1.800 & 3.92 & $7.50$ & 1130 & 20 & $14.96$ & 5193 \\[5pt] 
        \hline
    \end{tabular}
    \caption{Source parameters (redshift $z$ and spectral index $\Gamma_{\text{true}}$) of the synthetic sample described by a PL intrinsic spectrum and EBL attenuation according to the F10 model as implemented by \cite{2019pimentel}. For the ``short'' (left side) and ``long'' (right side) observation times (Obs. T., in hours), we show their respective TS values and the maximum energy bin of their spectra.}
    \label{tab:sources}
\end{table}

\begin{figure}[tbp]
    \centering
    \includegraphics[width=\textwidth]{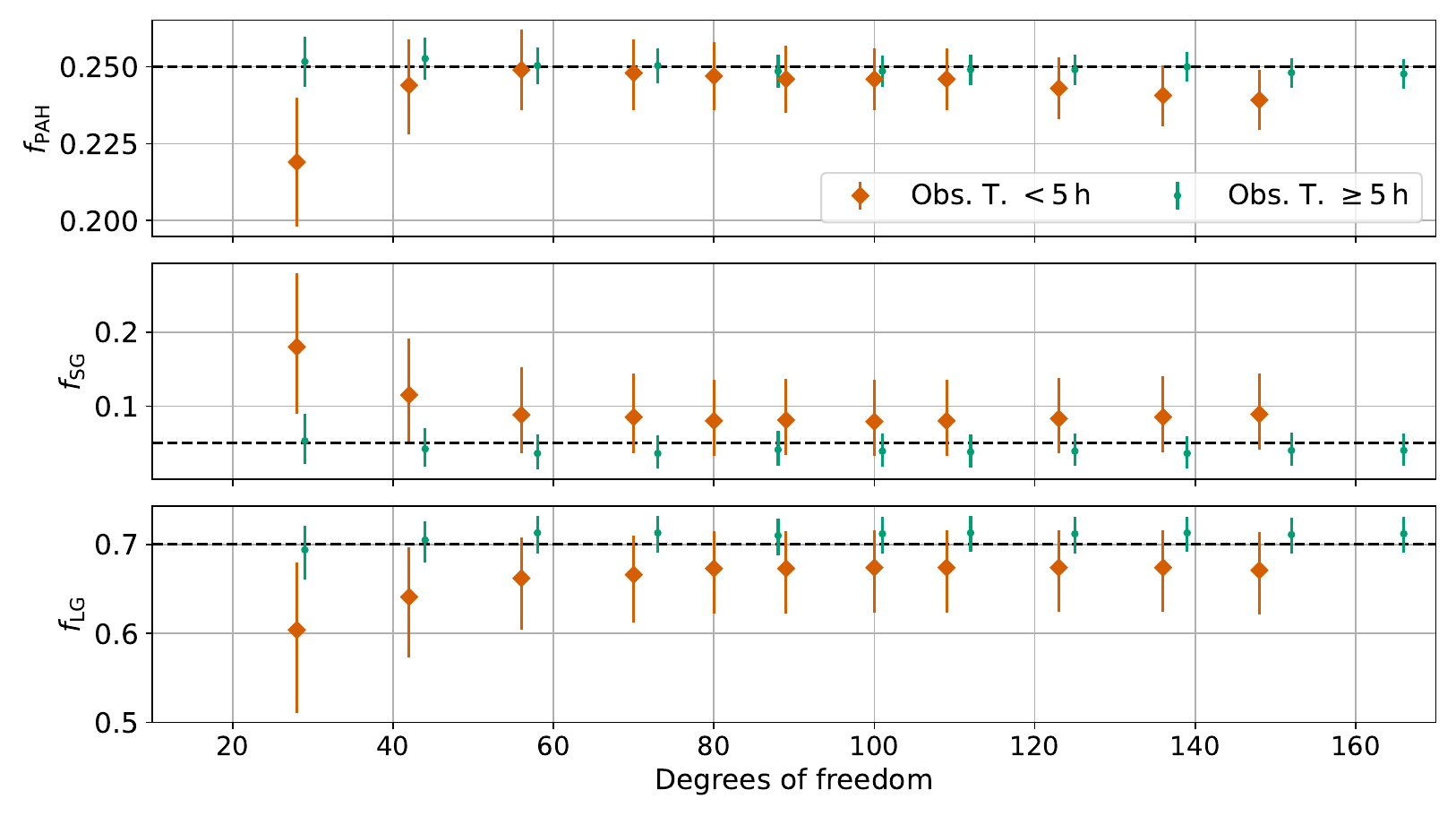}
    \caption{Median value of the EBL parameters (dust fractions) for the simultaneous fit from two to twelve synthetic sources with the highest TS, plotted as a function of the number of degrees of freedom. The error bars represent the 68\% interval of the marginal posterior distribution centered around the median. Vermilion diamond points are simulations with a total observation time lower than $5$\,h, while the bluish-green dots use spectra of sources observed for five or more hours. The black dotted lines are the true values used to compute the EBL attenuation.}
    \label{fig:paramsevolution_PLF10}
\end{figure}

\begin{figure}[tbp]
    \centering
    \includegraphics[width=\textwidth]{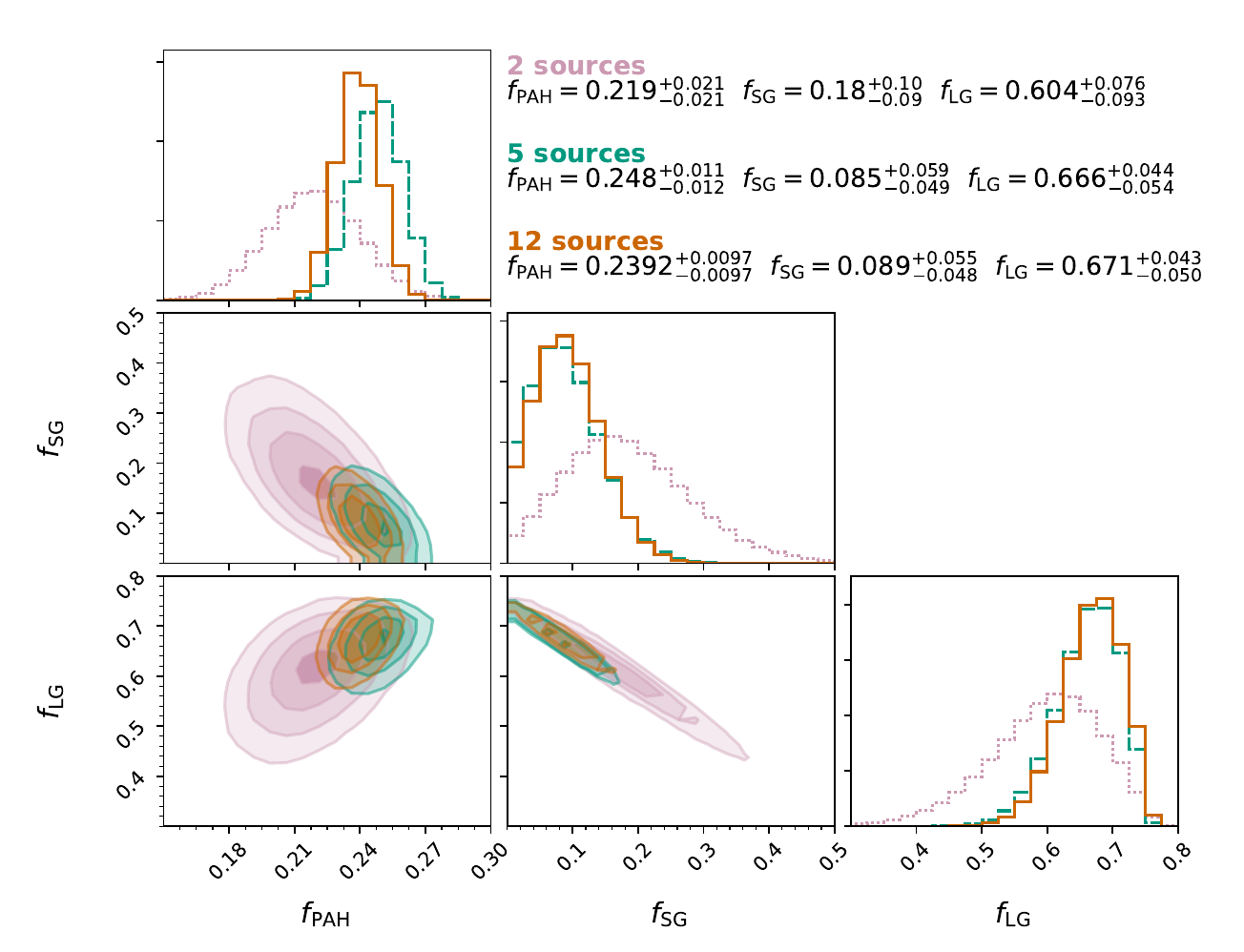}
    \caption{Synthetic sample posterior marginal distributions of the EBL parameters (dust fractions) in their 1-dimensional and 2-dimensional projections for observation times below 5\,h. The reddish-purple dotted histogram corresponds to the 2 sources fit, while the dashed bluish green to 5 sources and the vermilion solid line to 12 sources. The median values and respective uncertainties are indicated on the top right.}
    \label{fig:corner_PLF10}
\end{figure}

Figure~\ref{fig:paramsevolution_PLF10} presents the evolution of the median value of dust fractions, extracted from their marginal distributions, considering the spectra of short and long observation time (vermilion diamonds and bluish-green dots, respectively). Each point corresponds to a simulation with a given number of sources, presented in terms of the number of degrees of freedom (ndof). Since each simulation adds a new spectrum as input, the leftmost point is the case with the two highest TS sources, while the rightmost is the result for 12 sources. The error bars represent the 16th and 84th percentiles of each distribution, as to closely match the $1\sigma$ interval in the case of a Gaussian profile (68\% credible interval). We see, in general, an approximate convergence to the true values (dotted lines) as we increase the number of sources and degrees of freedom of the combined fit, as expected in the absence of systematic errors. Considering the data for low observation time, the two highest TS sources ($\text{ndof}=28$) can loosely describe the correct EBL levels, but the addition of new sources (with sequentially lower TS) reaches some saturation around five sources ($\text{ndof}=70$), beyond which the median value and the uncertainty interval don't substantially change compared to the simulation with 12 sources, especially for the SG and LG fractions. This behavior is not only present in the median but also in the whole shape of the marginal distributions, as figure~\ref{fig:corner_PLF10} illustrates. This is more pronounced for the SG and LG distributions, as it seems to indicate that constraints to the far-IR EBL are limited by the available spectra.

For the short observation time, the PAH fraction is the best constrained parameter, since its uncertainty represents a smaller percentage of the median, compared to the values for the SG and LG fractions, which have much broader distributions, as revealed by figure~\ref{fig:corner_PLF10}. However, the median values of dust fractions (which closely match the mode in this case) are systematically above or below the true parameters. We attribute this difference to the measured VHE photons being unable to properly disentangle the SG and LG components, as there is a strong negative correlation between them (of $\rho_\text{SG-LG}=-0.98$ for 12 sources). Even though the dust components are linearly related as $f_{\text{LG}} = 1 - f_{\text{PAH}} - f_{\text{SG}}$, the 2-dimensional marginal distribution between PAH and SG result in a weaker correlation ($\rho_\text{PAH-SG}=-0.52$), while for PAH and LG components we find a positive one, of $\rho_\text{PAH-LG}=+0.36$. Indeed, only when better spectral data are incorporated through longer observational periods (reaching higher energy bins and lower uncertainties) that we find a more accurate estimate of the SG and LG fractions, as figure~\ref{fig:paramsevolution_PLF10} shows. In fact, this is already apparent with two sources in the longer observational time ($\text{ndof}=29$), where the interval around the median is distant less than $1\sigma$ to the true value for all parameters. Once again, there seems to be a saturation of the resolution on dust fractions around 5 sources, with the addition of new spectra generating diminishing returns (gradually smaller uncertainties) beyond that.  Remarkably, the parameters in the two sources' case are better constrained than the 12 sources' results at low observation time, reinforcing the importance of having well-measured VHE spectra to constrain EBL parameters. 

\subsection{Probing systematic effects associated to the choice of EBL model}\label{sec:simul_D11}

The true EBL attenuation is not exactly known, but the available data coming from the counting of resolved sources and attempts of disentangling the EBL from foregrounds has provided lower and upper limits on its spectral intensity. 
At some wavelengths, however, the range of specific intensities allowed by limits and/or the lack of any data constraints, leaves room for a broad range of models to survive. 

Here, we test the impact of our choice of EBL model by preparing an input synthetic sample attenuated by the D11 EBL model \cite{dominguez2011extragalactic}, while fitting the dust fractions from the F10 model. This should allow us to explore to which extent a model based on stellar and three dust grain contributions is able to mimic the absorption of another widespread model like D11; and, more importantly, to quantify any possible systematic effects introduced by the wrongly chosen attenuation model into the reconstructed intrinsic source parameters. 

\begin{figure}[tbp]
    \centering
    \includegraphics[width=\textwidth]{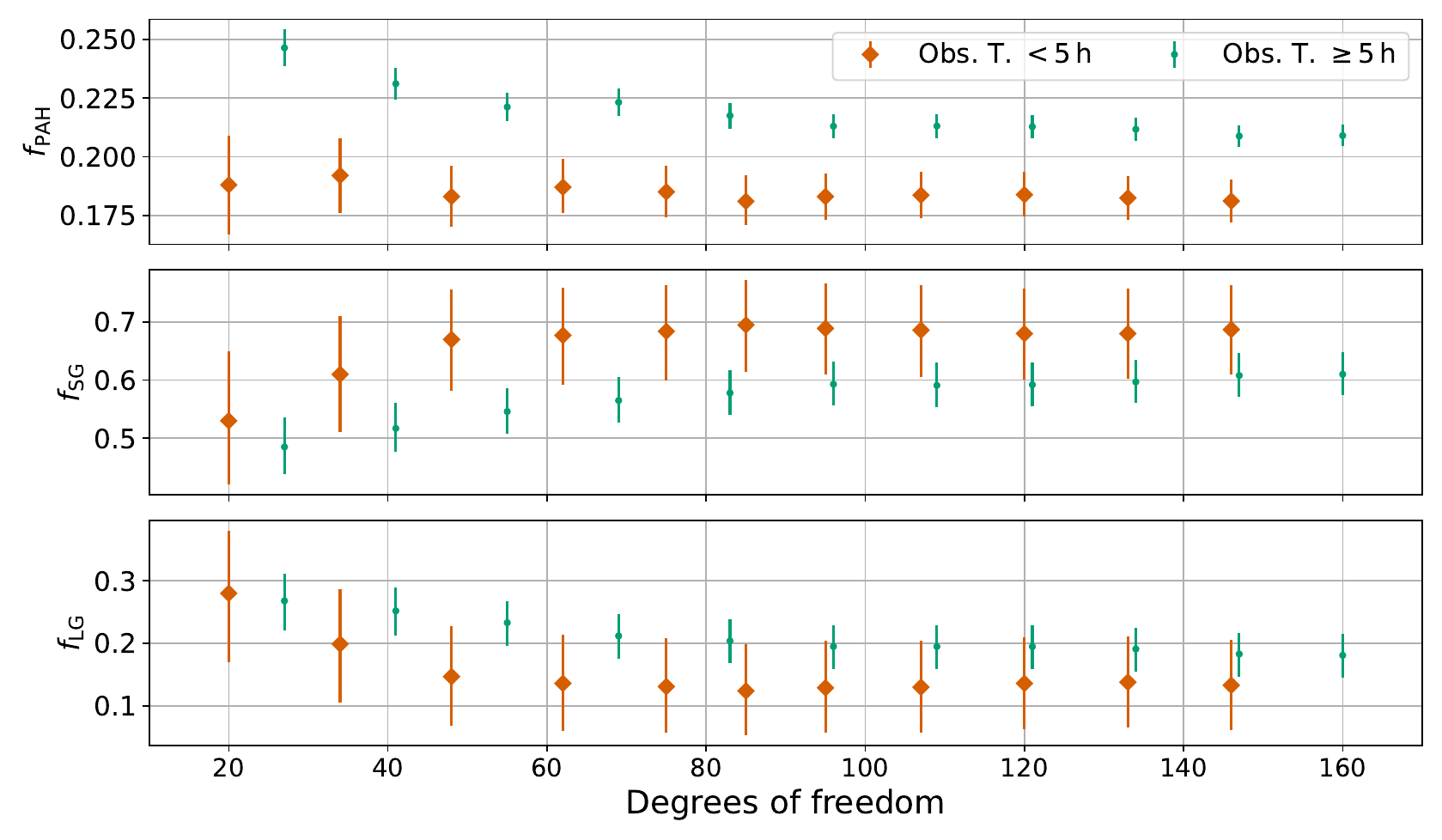}
    \caption{Synthetic sample median values of the EBL parameters (dust fractions) for the simultaneous fit from two to twelve sources with the highest TS (plotted as a function of the number of degrees of freedom), where a systematic effect has been introduced by attenuating the source fluxes with the D11 EBL model \cite{dominguez2011extragalactic}.}
    \label{fig:evol_params_D11}
\end{figure}

Thus, all previously simulated sources had their observed spectra recomputed adopting D11 model as the true EBL. Once again ordering the sources from highest to lowest TS, we have repeated the procedure of simultaneously fitting EBL and intrinsic parameters using the two distinct sets of total observational time. Figure~\ref{fig:evol_params_D11} shows the parameter evolution for the sequence of fits in an increasing number of degrees of freedom. As a consequence of the introduction of a systematic effect in the EBL model, a corresponding systematic shift is introduced into the recovered dust fractions. The biggest difference being the SG and LG components, as the higher EBL intensity in the far-IR for the D11 model compared to F10 and \cite{2019pimentel} (parameters from table~\ref{tab:true_values_ebl}) demands a larger abundance of the colder components.

The increment in observation time did not significantly change the inference of the EBL levels, but a few differences can be pointed out from figure \ref{fig:evol_params_D11}. For example, although the SG and LG fractions continue to agree under $1\sigma$ for both observational times, there is a noticeable shift in the PAH fraction, converging to a higher value (above $0.20$) compared to the case in which the total observation time is less than 5\,h (below $0.20$). This shift may be attributed to a systematic effect introduced through the incorrect EBL model. Since the F10 model with normalised free fractions has lower EBL intensities at far-IR wavelengths ($>\SI{60}{\micro\meter}$), when fluxes at high-energies ($E\gtrsim 10$ TeV) are clearly measured, as is the case for longer observation times, the fit tries to compensate the lack of SG and LG absorption by increasing the optical depth due to PAH.

\begin{figure}[tbp]
    \centering
    \includegraphics[width=\textwidth]{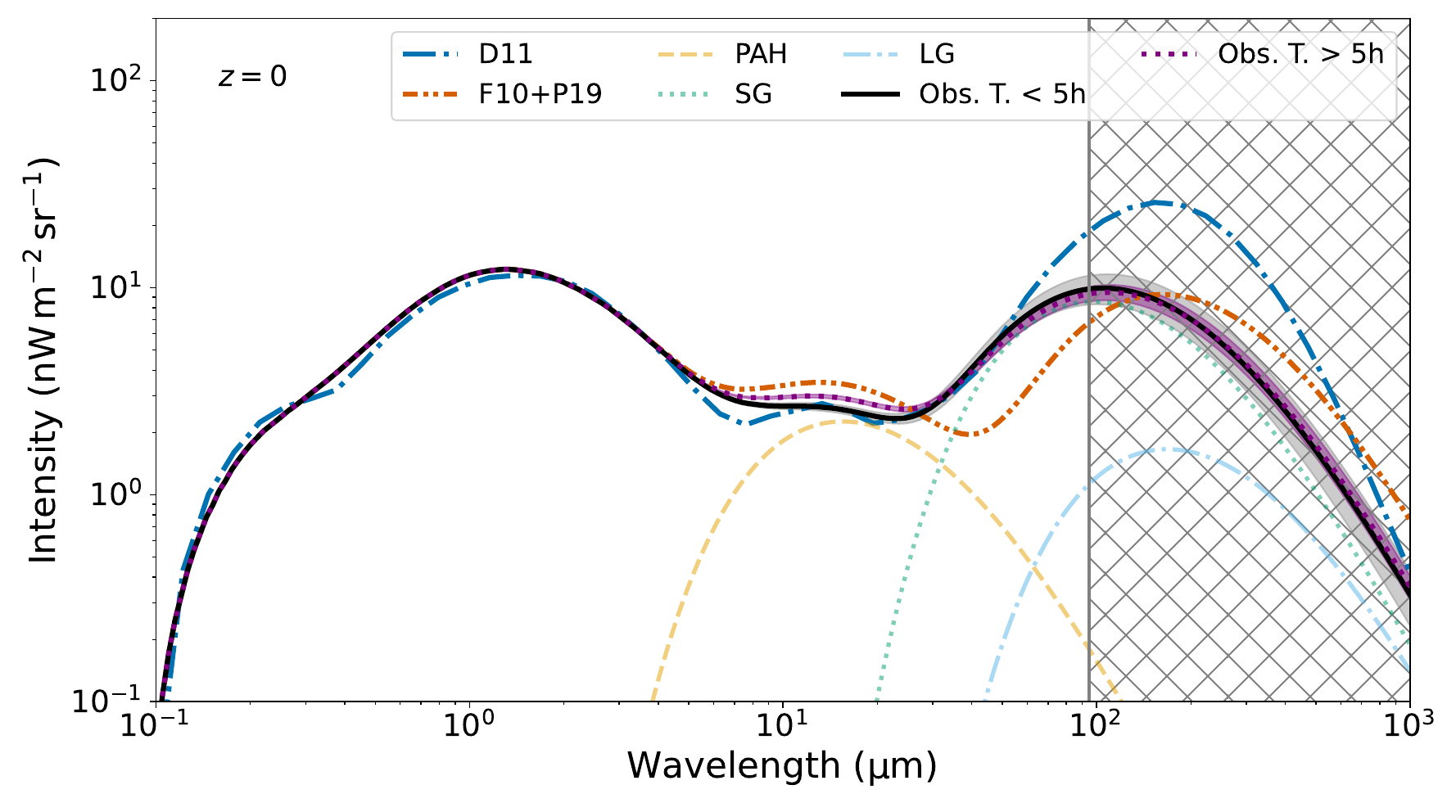}
    \caption{EBL intensity (black solid line) reconstructed from the inference of dust fraction using 12 spectra in the short and long observation time category. The contribution from each dust component for the short observation time is plotted separately as indicated by the legend. All sources had their intrinsic spectra attenuated by the D11 model (blue dash-dotted line) and we show for comparison the F10+P19 nominal model, consisting of the base values presented in table \ref{tab:true_values_ebl}. The hatched region corresponds to the EBL interval that does not interact with the gamma rays of the dataset.}
    \label{fig:ebl_D11}
\end{figure}

A comparison between the reconstructed EBL at $z=0$ and the true D11 shape can be seen in figure~\ref{fig:ebl_D11} and more clearly shows how effectively F10 model with free fractions can reproduce the D11 description. The solid line represents the EBL density obtained from the median values of the dust fractions --- considering the 12 sources inference with lower observation time --- and the $1\sigma$ uncertainties are represented by the shaded area around the curve. On one hand, F10 and D11 models have a good agreement between $\SI{0.1}{\micro\meter}$ and $\lesssim \SI{2}{\micro\meter}$, an EBL region dominated by stellar emission, which is fixed in our analysis. However, F10 predicts a larger mid-IR energy density and a lower far-IR one compared to D11. When the dust fractions are allowed to vary, we observe that the agreement between these two models can be extended beyond the stellar region, up to $\lesssim\SI{60}{\micro\meter}$. In the same figure, we also plot the contribution to the energy density from each dust component in the free fractions model. We see, particularly, the important role of the small grains to capture the rise in intensity above $\SI{20}{\micro\meter}$, while the PAH component is able to describe the correct EBL level in the region $\SI{10}-\SI{100}{\micro\meter}$. 

\begin{figure}[tbp]
    \centering
    \includegraphics[width=\textwidth]{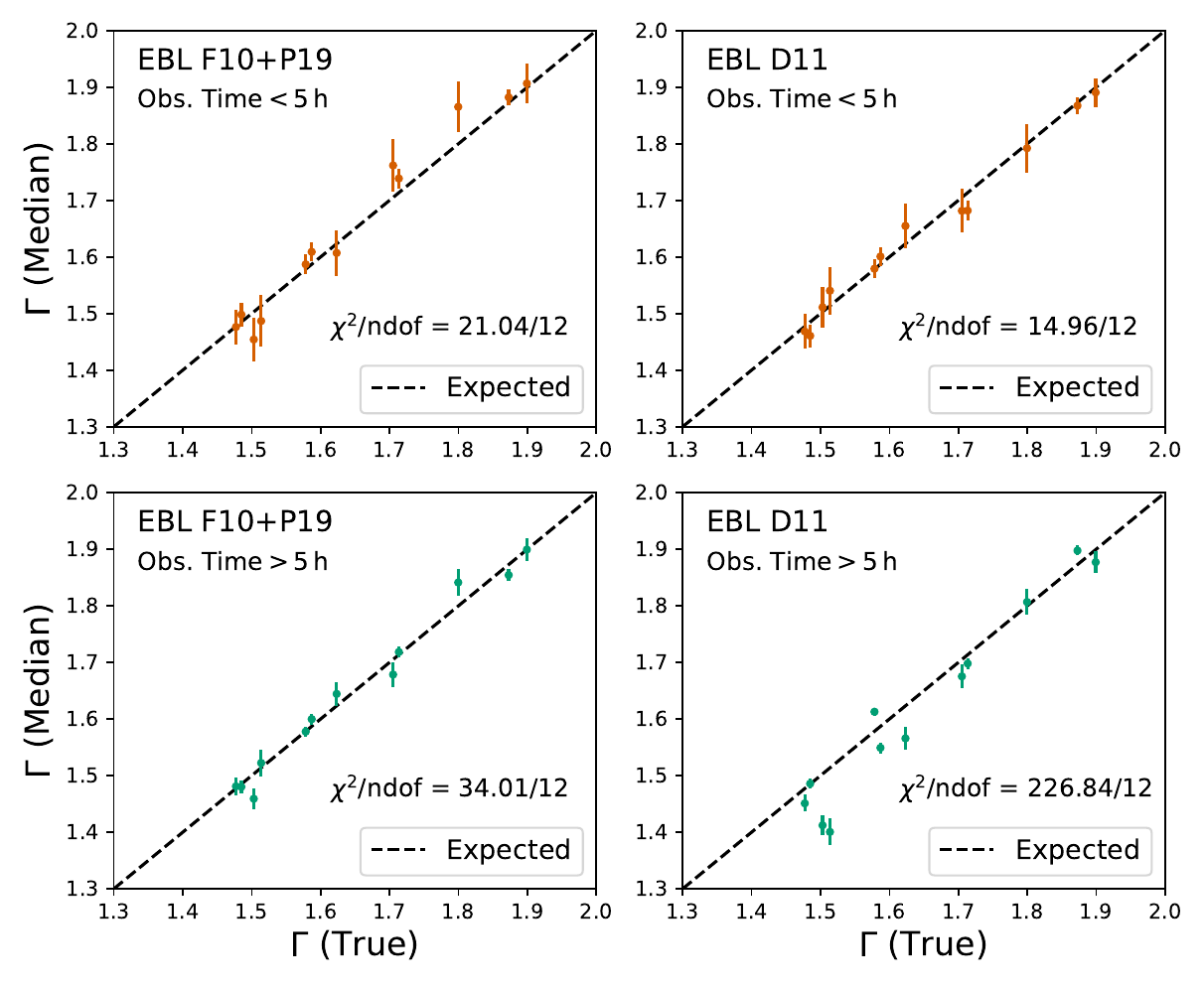}
    \caption{Comparison between the median value of spectral indices and their respective true value when sources were attenuated by F10+19 EBL model (left column) and D11 (right). Each subplot contains the result of a simulation using 12 sources (short and long observation periods in top and bottom rows, respectively). The chi-square values between the data (median) and expected values (dashed line) are also displayed.}
    \label{fig:specindex_D11}
\end{figure}

Figure~\ref{fig:specindex_D11} compares the median spectral indices from the combined fit of 12 sources with their respective true values, again for two scenarios, depending on whether or not the likelihood EBL matches the true model used to attenuate the synthetic SEDs. We can see that for observation times below 5h, there are no hints, from the statistical point of view, of the presence of systematic biases introduced in the reconstructed intrinsic spectral indices. On the other hand, when sources are observed for more than 5h, a few outliers are present in the bottom-right plot. The reduced $\chi^2$ (226.4/12), calculated with respect to the expectation (represented by the diagonal dotted line), indicates that the observed scatter is not statistically consistent with the estimated uncertainty coming from the fit. It is also clear from the bottom right plot that the bias is more pronounced for harder spectra. This is expected because, for these sources, the intrinsic SEDs extend to higher energies, probing in turn, the SG- and LG-dominated parts of the EBL spectrum. As we can see from figure~\ref{fig:ebl_D11}, in order to mimic D11 attenuation, the fit will tend to reduce slightly the PAH contribution at mid-IR and compensate for part of the attenuation generated by these molecules' emissions by overestimating the SG and LG contributions. In the end, a residual bias survives, with the fits associated with harder sources converging to underestimated values of photon indices in order to balance the overestimated SG and LG contributions at the far-IR ($\gtrsim\SI{100}{\micro\meter}$).

\begin{figure}[tbp]
    \centering
    \includegraphics[width=\textwidth]{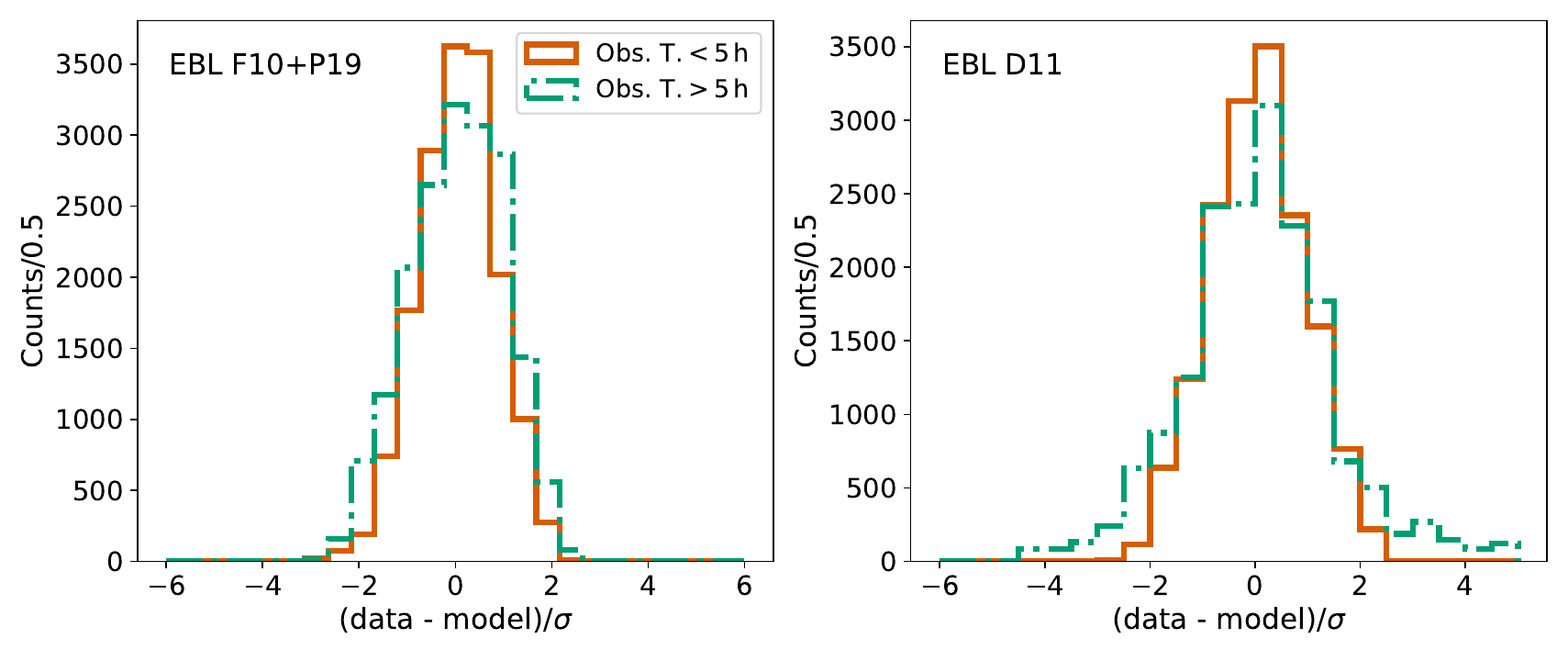}
    \caption{Residuals of the fit between all the spectra of 12 highest TS sources and their respective models. The parameter values of each model were extracted randomly (100 times) from the posterior distribution and we performed a maximum likelihood fit to obtain the flux normalization for each source. The histograms, then, present the resulting counts (over the bin interval). The dot-dashed lines are the results for total observation time greater than $5$\,h, while the solid lines present the results for observation time lower than $5$\,h.}
    \label{fig:resid_histogram}
\end{figure}

Next, we turn to the question of the SEDs fit quality. In other words, if the EBL and intrinsic parameters inferred act, together, as a good description of the observed spectra. Since the intrinsic SED normalisation has been marginalised during the posterior sampling, we now sample random points from the posterior, taking these parameter values as input to a maximum likelihood fit of the flux normalisation only. Therefore, for each source, we can recover its intrinsic spectrum, eq.~\eqref{eq:intrinsic}, and analyse the fit quality from the distribution of the residuals. This procedure can probe inconsistencies or nonphysical spectra that may emerge from a poor combination of parameters. The random selection of models from the chain also helps us to identify if parameters from the high-density regions of the posterior distribution can be a valid description of the unknown EBL and intrinsic parameters, given all the information available. Taking the posterior distributions for the 12 sources in the scenarios presented before (with and without systematic errors), we proceed to make the described fit for each spectrum and then, construct histograms of the residuals to identify outliers. Figure \ref{fig:resid_histogram} presents the pull distribution of 100 random models from the posterior, superimposing results of different observation times for comparison. The most distinguishable feature in all histograms is the presence of long tails when there is a mismatch between the EBL model used during the generation of the synthetic sample and that for the construction of the likelihood. However, the absence of this feature for observation times lower than $5$\,h reveals that only when sufficiently well-measured spectra are available that systematic effects associated with the modeling of the EBL density will be properly probed.

\section{EBL and intrinsic parameters constraints from IACT data}\label{sec:tevcat}

Given all the spectra of the sources described in table~\ref{tab:tevcat_sources}, we need to choose the intrinsic model that best suits each one. The analysis performed by \cite{2015ApJBiteau} reveals that most gamma-ray sources observed by ground observatories until 2014 can be well described by the power-law model, but some spectra show signs of intrinsic curvature that are compatible with the LP or PLC functions. Our choice of intrinsic spectrum parametrization was based on an exploratory sampling of the posterior distribution using as input data the individual SEDs. That is, we have run MCMCs for each source individually and analysed the general features of their posterior distributions when different intrinsic spectrum assumptions were made. Special attention was paid to posteriors where the grain fraction values differed radically from the nominal F10 values.

For example, some input SEDs (like those of Markarian 421 and some of PKS 2155-304) resulted in exceedingly high PAH fractions ($f_{\text{PAH}} \gtrsim 0.70$) when PL was chosen as the intrinsic spectrum. This could be explained if there was an intrinsic curvature requiring higher attenuation levels at the TeV range to compensate for the lack of curvature in the fitting model. To verify it, the one source MCMC simulations for these spectra were repeated using LP and PLC models, which indeed lowered the required EBL levels. On many occasions, however, the PLC model is difficult to sample, as a strong degeneracy in the cut-off appears, demanding an ad hoc upper limit on the range of $E_\textrm{cut}$ values to produce a reasonable posterior distribution. Therefore, in these cases, we adopted the LP model. We also searched the literature for any evidence of intrinsic curvature in previous studies.

In fact, a recurrent case is Markarian 421, as \cite{2007Albert} concludes that the curvature in the MAGIC 2004-2005 data is partially source-inherent and the intrinsic spectrum is compatible with PLC. The same spectral model is also used by \cite{CoutinodeLeon:2017cse} to describe HAWC 2015-2017 data in the presence of EBL attenuation, whereas \cite{2011mrk421} fits VERITAS 2008 data, although not including EBL effects. Another noteworthy case is PKS 2155-304 data from HESS 2006, as analyzed by \cite{HESS:2013udx} in the context of axion-like particle constraints, in which an LP model is used to describe the intrinsic spectrum. Table \ref{tab:tevcat_results_curvature} presents the data for which the PLC or LP models were used during our analysis, whereas tables \ref{tab:tevcat_results}, \ref{tab:tevcat_results2} and \ref{tab:tevcat_results3} summarize the PL spectra\footnote{Differently from \cite{2015ApJBiteau}, we adopted the PL model for 1ES 2344+514 (VERITAS/2007-2008), PKS 2155-304 (MAGIC/2006), and Markarian 421 (VERITAS/2008 - high C), instead of a LP model. Also, we adopted the PLC in place of LP (and vice-versa) for Markarian 421 VERITAS/2008 very low, low, high A, high B, and very high data.}.
\begin{figure}[t]
    \centering
    \includegraphics[width=\textwidth]{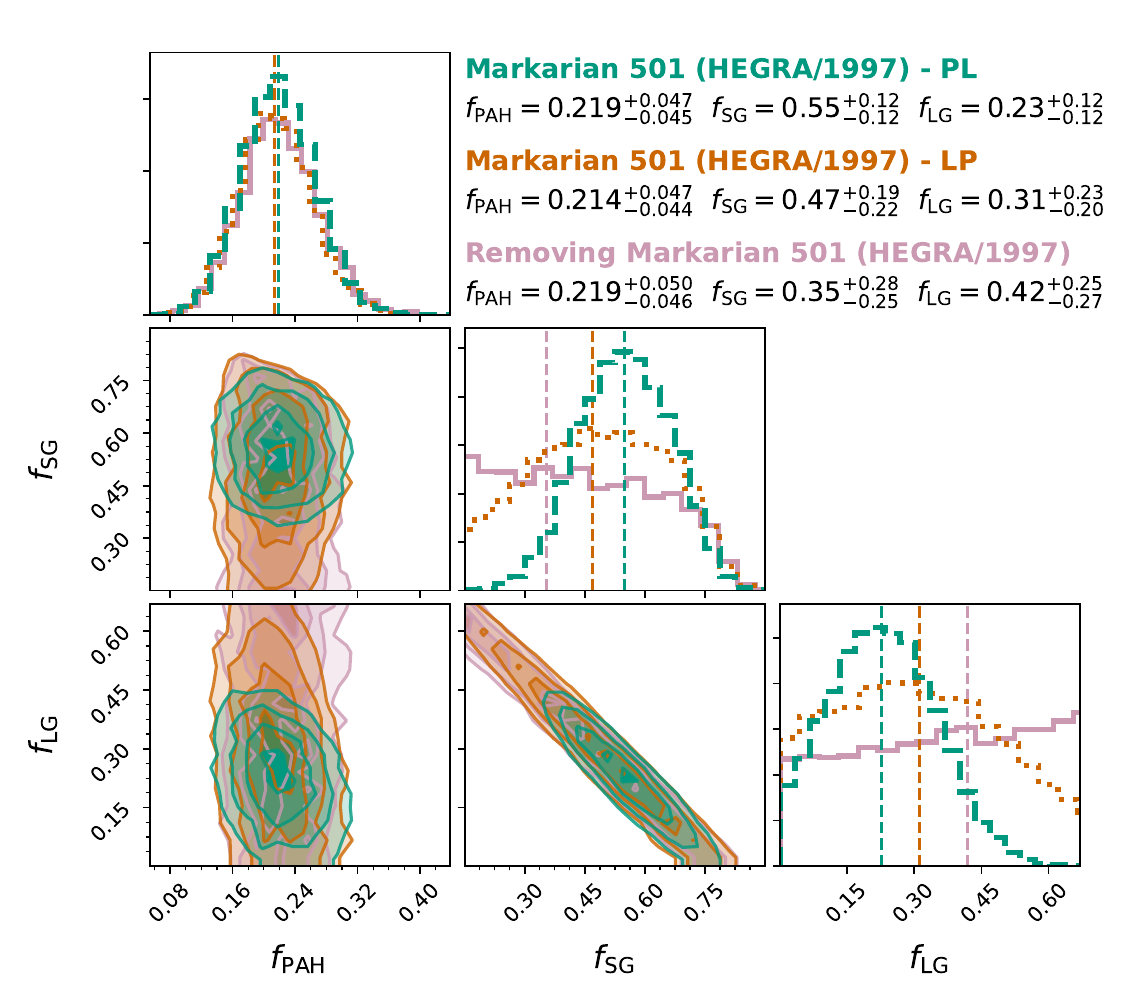}
    \caption{Marginal distributions of EBL dust fractions for the three cases using IACT data (including Mkn 501 flare spectrum as a power law and log parabola, or excluding this data from the likelihood). For each parameter, the median of the 1D-distribution is displayed as the vertical dashed line.}
    \label{fig:corner_tevcat}
\end{figure}

Particular attention had to be paid when choosing the intrinsic model for Mkn 501 (HEGRA/1997, flare state), since, as discussed by \cite{2019pimentel}, it has the highest expected attenuation by the dust component of the EBL (relative to the stellar component). Besides, depending on the choice of spectral model, the fitted EBL parameters can change considerably. With this in mind, we ran HMCs with three different configurations: a) excluding Mkn 501 flare SED from the sample; b) including it together with a PL model, and c) using the LP parametrization.

The posterior distribution for EBL parameters can be seen in figure \ref{fig:corner_tevcat}. As far as the current VHE SEDs are concerned,  Mkn 501 SED is essential for constraining the far-IR portion of the EBL spectrum. In fact, the combined information of all 64 other spectra used in this analysis resulted in mostly flat marginal distributions for these colder dust components. Moreover, the constraints on SG and LG fractions are stronger in the absence of intrinsic curvature (PL model) for Mkn 501 flare SED. Naturally, if the flux levels at the highest energies of Mkn 501 can be described by some intrinsic curvature, the extra parameter will introduce a degeneracy with respect to the EBL attenuation, widening the marginal posterior distribution of the corresponding dust components. Finally, the PAH component is constrained by the combined inference of all sources, as the removal of Mkn 501 or the change of intrinsic model only marginally impacted this dust fraction. This reinforces the important prospects of using combined IACT data -- even current measurements -- to put constraints in the mid-IR portion of the EBL.

Compared to \cite{2019pimentel}, where only Mkn 501 (HEGRA/1997) was used to find best-fit values for the dust fractions, the combined fit of this work reduces the uncertainty in the PAH fraction, regardless of the intrinsic model chosen. Specifically, when the intrinsic spectrum of Mkn 501 is described by a PL, reference \cite{2019pimentel} found $f_{\text{PAH}}=0.32\pm0.15$, which is consistent with the current result, although with an uncertainty approximately three times higher. Similarly, when adopting the LP model, \cite{2019pimentel} found $f_\text{PAH}=0.27\pm0.25$, an uncertainty approximately five times higher. Clearly, the combined use of other sources in the likelihood was key for improving the precision of EBL parameters. However, for the SG fraction (constrained essentially by Mkn 501), we have obtained approximately the same result as \cite{2019pimentel} ($f_{\text{SG}}=0.56\pm0.12$), including its uncertainty. In the LP scenario, it also agrees with the best-fit values of $f_{\text{SG}}=0.49\pm0.28$, although the uncertainty in \cite{2019pimentel} is a bit larger.

\begin{figure}[t]
    \centering
    \includegraphics[width=\textwidth]{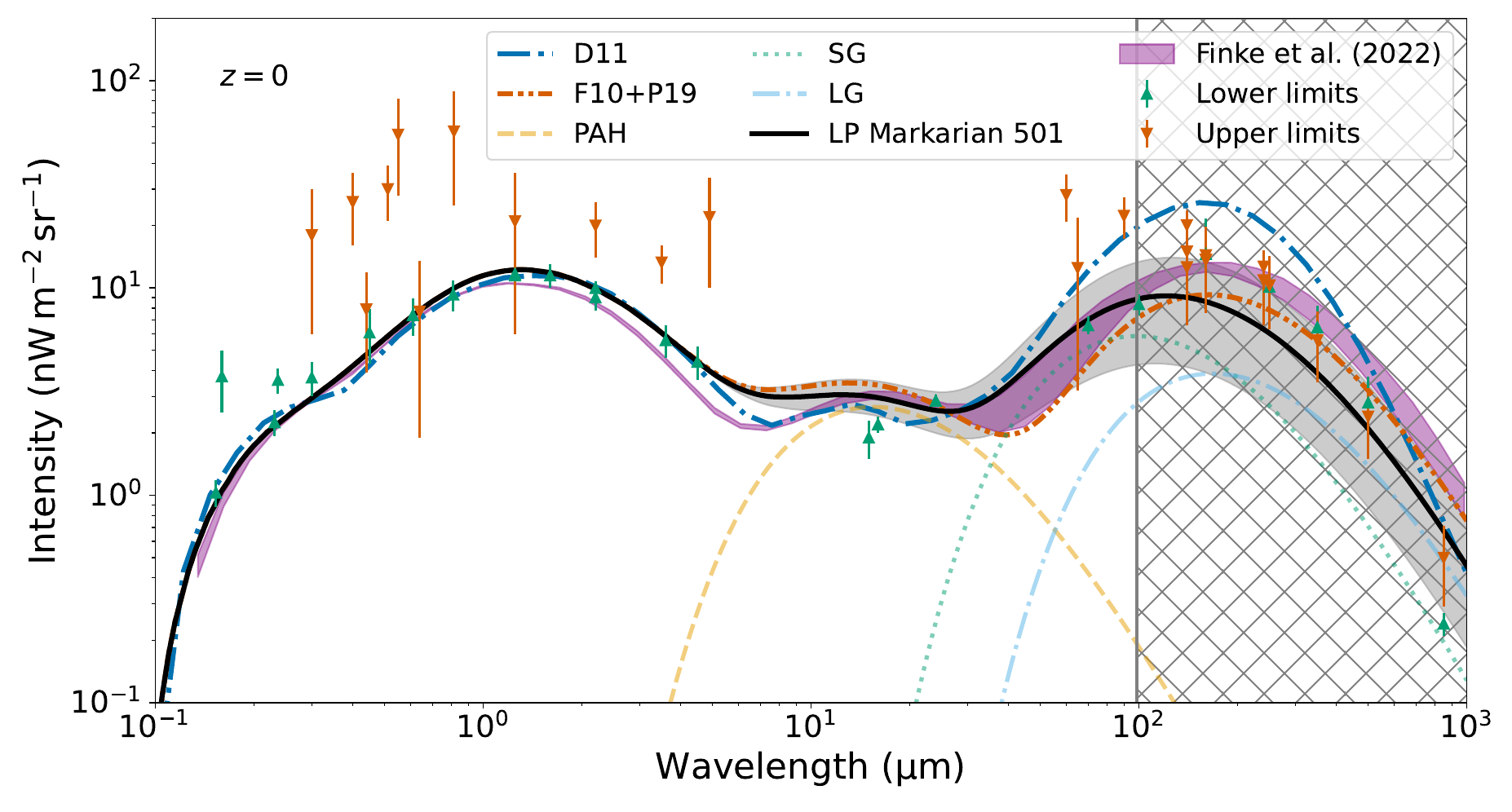}
    \caption{Local EBL intensity inferred using IACT data (solid black line with gray band) and adopting the LP model to describe Mkn 501 flare spectrum. The shaded gray region represents the lower and upper uncertainties of dust fractions. The individual contributions of each dust component are also shown (PAH, SG, LG), alongside other models. The upper and lower limits of EBL measurements were extracted from \cite{2015ApJBiteau}. The hatched region corresponds to the EBL interval that does not interact with the gamma rays from the dataset.}
    \label{fig:ebl_spec_tevcat}
\end{figure}

Figure \ref{fig:ebl_spec_tevcat} shows the corresponding local EBL spectrum when Mkn 501 is included in the combined fit with an LP intrinsic spectrum. The PL case result is very similar, with a slightly higher SG contribution and narrower uncertainty bands. Compared to the F10 and D11 models, up until $\sim$\,$\SI{100}{\micro\meter}$, the current result lies between these two. Notably, around $\sim$\,$\SI{30}{\micro\meter}$, the inferred EBL level is very close to D11's curve. However, differently than what is presented by \cite{2019pimentel}, which demands $f_\text{SG}=0.05$, we find a much higher SG fraction. 

Recently, an updated EBL model based on F10 was published \cite{2022ApJFinke} (Finke et al. 2022) and it is also included in figure \ref{fig:ebl_spec_tevcat}. It takes improved stellar models, a different parametrization of the star formation rate, and incorporates metallicity and dust extinction evolution with redshift. The authors utilise a variety of data (EBL opacity from LAT and IACTs; luminosity, stellar density, and dust extinction from galaxy surveys) to fit their model parameters, including the dust fractions. As a result, the model predicts $f_\text{SG}=0.26^{+0.18}_{-0.17}$ and $f_\text{LG}=0.56^{+0.17}_{-0.18}$, however, the LG dust temperature is also a free parameter in this model, leading to a value $T_\text{LG}=60.5^{+2.3}_{-3.5}\,\si{\kelvin}$, i.e., higher than F10, where $T_\text{LG}=\SI{40}{\kelvin}$ was fixed. The fitted LG temperature is closer to the SG value of $T_\text{LG}=\SI{70}{\kelvin}$. The local EBL spectrum is consistent with our result in the range $10$--$\SI{100}{\micro\meter}$, with some differences seen in the near-IR to optical.

\begin{figure}[t]
    \centering
    \includegraphics[width=\textwidth]{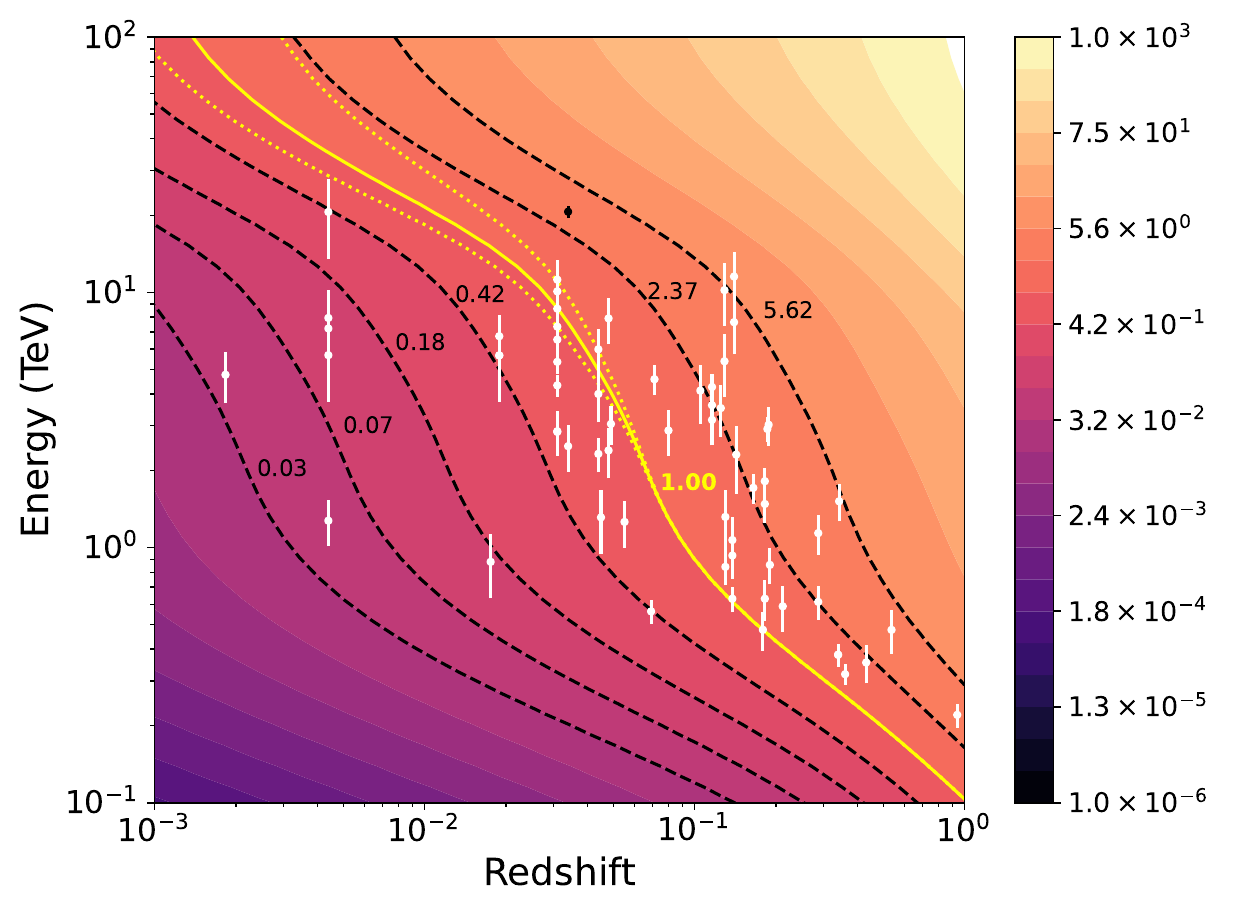}
    \caption{Optical depth map of EBL attenuation in the gamma-ray energy and source redshift plane. The optical depth is tuned to the median values of dust fractions from the LP Mkn 501 case (see figure \ref{fig:corner_tevcat}). Some level curves are displayed, with emphasis on the CGRH ($\tau=1$) in a yellow solid line. The dotted lines around it represent the positive and negative uncertainties of the median values. The white points are the last energy bins of every IACT spectrum used in this work and Mkn 501 (HEGRA/1997) is highlighted in black.}
    \label{fig:cgrh_tevcat}
\end{figure}

Figure \ref{fig:cgrh_tevcat} presents the optical depth in the redshift-energy plane, emphasizing the curve for which $\tau = 1$, called the Cosmic Gamma-ray Horizon (CGRH). Around it, we also consider the upper and lower uncertainties in the dust fractions, with the corresponding CGRH in these cases represented by yellow dotted lines. While the stellar component of the EBL density essentially determines the high-redshift portion of the CGRH, the IR (dust-dominated) portion of the EBL is more directly probed by local ($z < 0.1$) strong TeV emitters (reaching energies of a few dozen TeV). In fact, the Mkn 501 spectrum (at $z = 0.034$) has the highest energy bin of our sample, above $\SI{20}{\tera\electronvolt}$ and was essential for constraining the far-IR region of the EBL. We see a concentration of sources near the CGRH, although some of them, with $z \lesssim 0.2$, are subject to stronger attenuation, reaching optical depths of $\tau = 5$ (where $e^{-5}\approx 0.0067$) or higher. One should be careful to avoid drawing statistical conclusions from figure \ref{fig:cgrh_tevcat} based on the presence of a non-negligible population of blazars at a region of high-opacity of the extragalactic medium. The sources used in this plot are a compilation of observed TeV spectra and are not coming from an uniform sky survey, therefore this sampling has some significant selection bias.   

\begin{figure}[t]
    \centering
    \includegraphics[width=\textwidth]{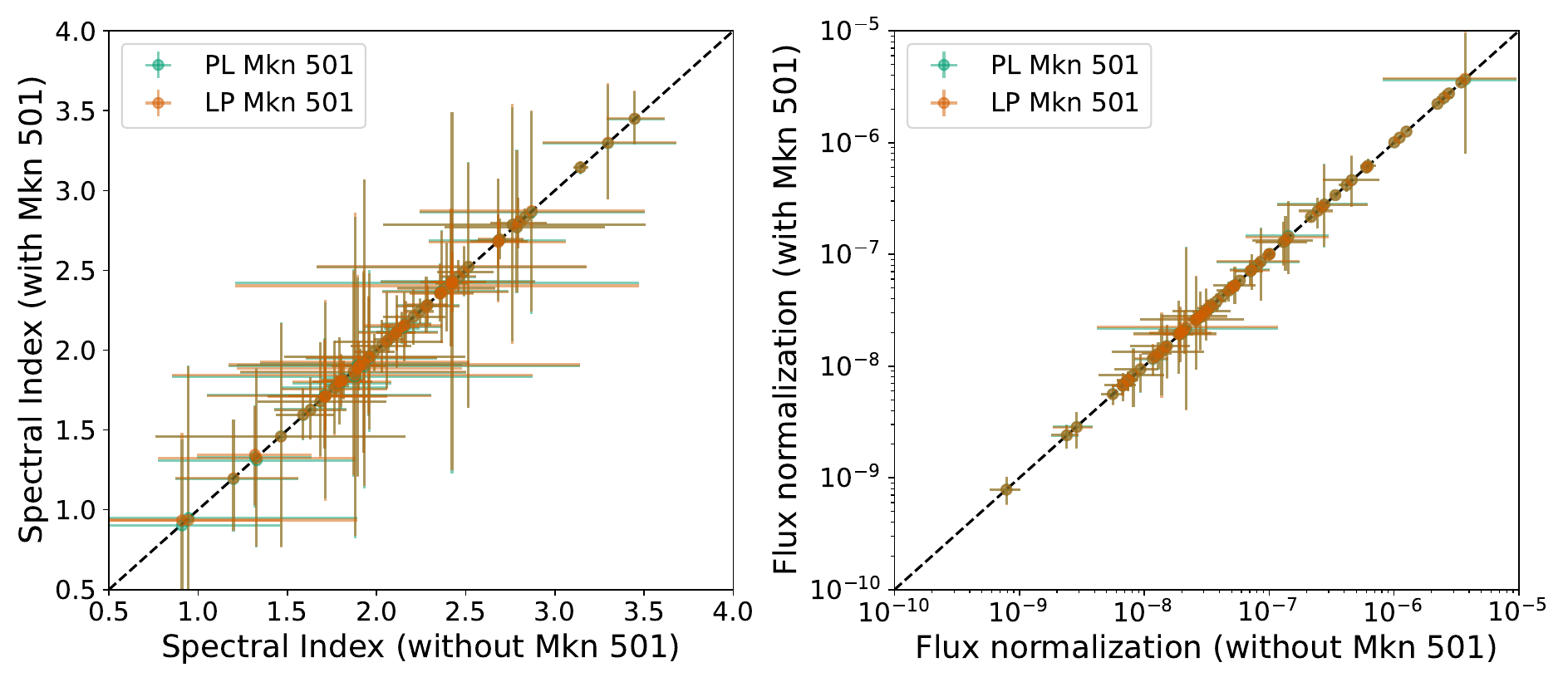}
    \caption{Comparison between fits with Mkn 501 (as PL or LP) and without it. \textit{Left}: median spectral index of the sources (including LP and PLC sources); \textit{Right}: flux normalization parameter. The black dashed line is the identity.}
    \label{fig:specindex_mkn501}
\end{figure}

As the statistical uncertainties in the EBL incorporate those of the intrinsic parameters, due to the marginalization of the posterior, the same is true, in turn, for the intrinsic parameters. However, since most of the sources cannot constrain the SG or LG fractions, their intrinsic parameters are generally insensitive to changes in the EBL in this region. Figure \ref{fig:specindex_mkn501} shows that there are no substantial differences in the median values and respective uncertainties of spectral indices and SED normalization when we change the model of Mkn 501 or remove the data from the sample. This result also strengthens our confidence in the robustness of the PAH constraint.

\begin{figure}[t]
    \centering
    \includegraphics[width=\textwidth]{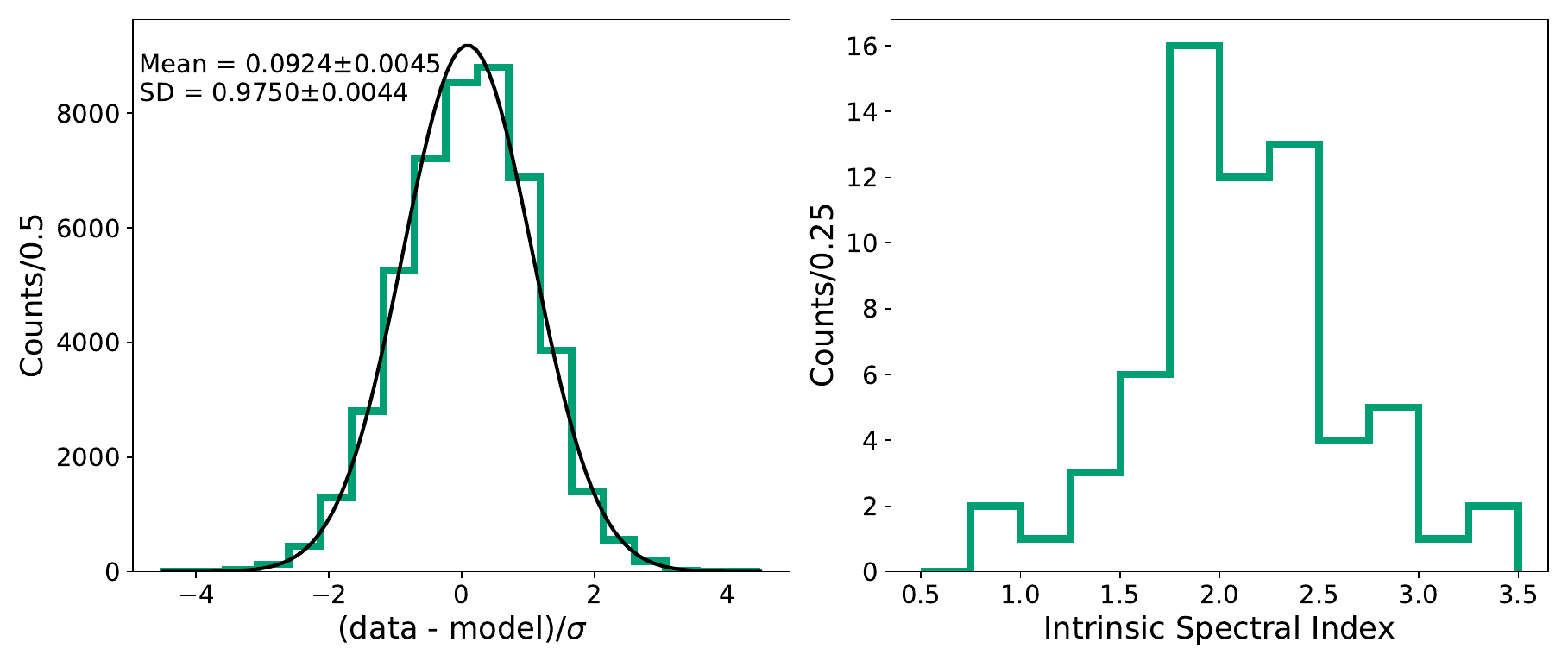}
    \caption{\textit{Left}: residuals of the fit of all IACT data used in this work (case where Mkn 501 was described with a LP model), from a sample of 100 random points of the posterior distribution. The solid line is the best-fit Gaussian curve to the histogram, using the unbinned data, with its mean and standard deviation written on the side. \textit{Right}: median spectral indices of the sources, including $a$ and $\Gamma$ from LP and PLC models.}
    \label{fig:resids_tevcat}
\end{figure}

Similar to what has been done for the synthetic sources, to better evaluate the adequacy of the intrinsic parameters, once the EBL attenuation is taken into account, we select a sample of 100 points from their posterior and computed the pulls (i.e. the residuals in units of standard deviations) between the observed data and the model prediction. This is shown as the histogram on the left-hand side of figure \ref{fig:resids_tevcat}. Performing an unbinned log-likelihood fit we found a standard deviation of $0.9750\pm 0.0044$ and a mean of $0.0924\pm 0.0045$. A complete consistency with a zero-mean and unit variance Gaussian distribution of pulls is not expected given the effective nature of the parametric intrinsic spectra adopted for TeVCat sources. Such models should be seen only as first approximations of the true intrinsic SEDs. Moreover, some deviation from a Gaussian distribution for the pull values is expected in the tails of the SEDs because at these spectral regions, the fluctuations are intrinsically Poissonian rather than Gaussian. The plot on the right-hand side of figure \ref{fig:resids_tevcat} presents the distribution of median intrinsic spectral indices. In general, most of the sources in the sample have a spectral index near $\Gamma \approx 2$ and below $2.5$, with the maximum median value being $3.45$ and the minimum $0.90$. In fact, six spectra resulted in $\Gamma < 1.5$: 1ES 0229+200 (HESS/2005-2006), 1ES 0440+009 (VERITAS/2008-2011), 3C 279 (MAGIC/2008), H1426+428 (HEGRA/2002), PKS 0447-439 (HESS/2009) and RBS 0413 (VERITAS/2009), although only H1426+428, with $\Gamma = 0.90^{+0.53}_{-0.50}$ has more than $1\sigma$ difference from $\Gamma=1.5$. Nonetheless, the result still agrees with \cite{2015ApJBiteau}, which has found $\Gamma=1.37\pm0.30$ for this source. Tables \ref{tab:tevcat_results_curvature}, \ref{tab:tevcat_results}, \ref{tab:tevcat_results2} and \ref{tab:tevcat_results3} present the median values and respective uncertainties for all parameters and sources.

\begin{figure}[t]
\centering
\begin{subfigure}{.5\textwidth}
  \centering
  \includegraphics[width=\textwidth]{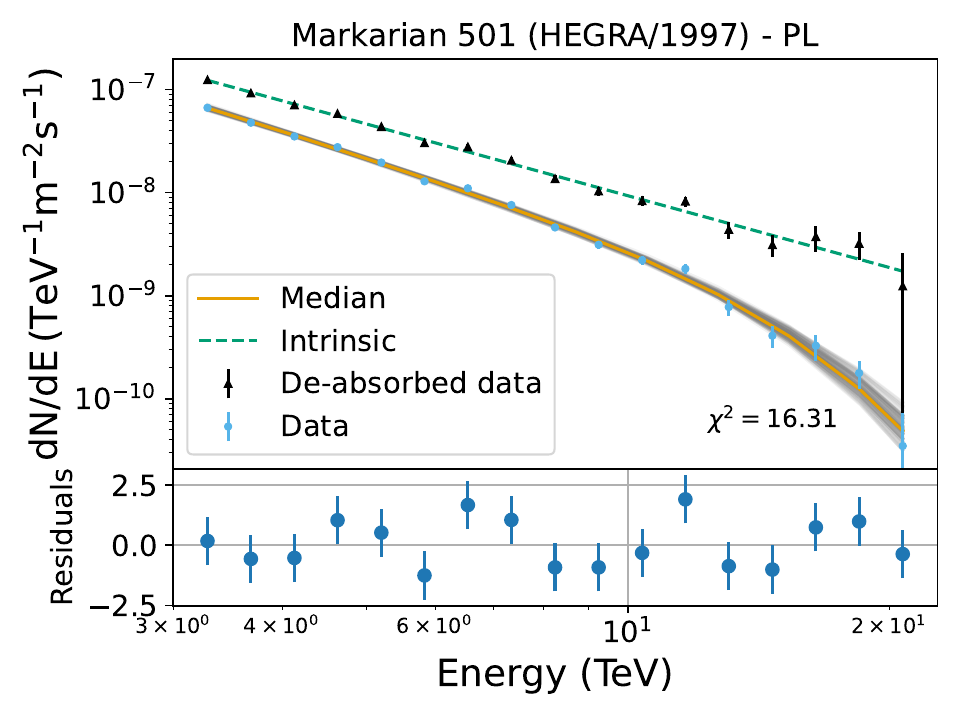}
\end{subfigure}%
\begin{subfigure}{.5\textwidth}
  \centering
  \includegraphics[width=\textwidth]{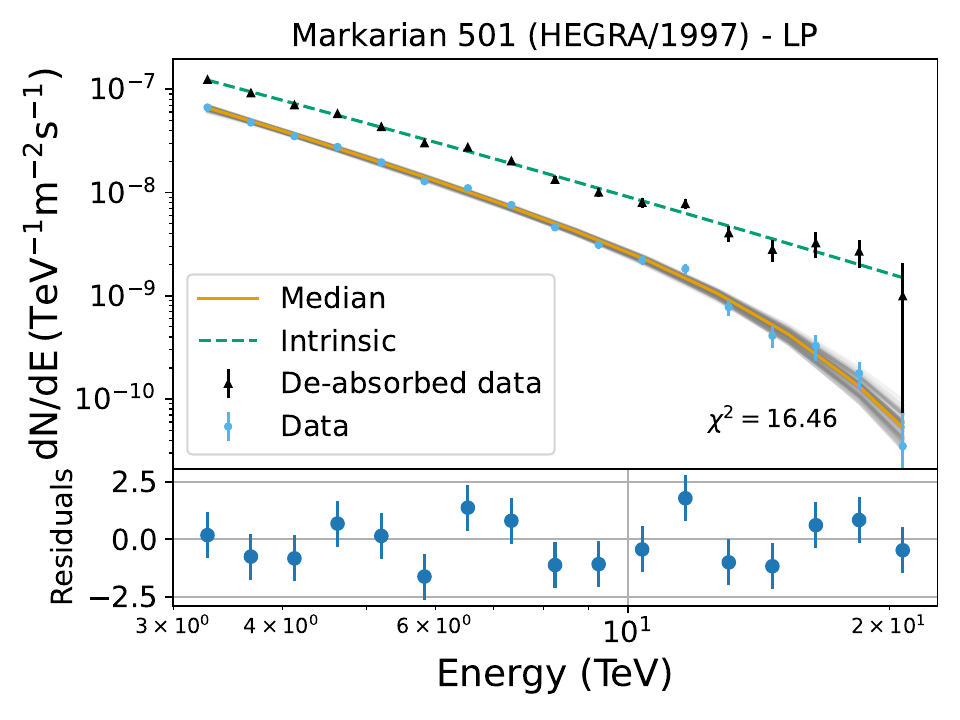}
\end{subfigure}
\caption{\small{Fit of Mkn 501 (HEGRA/1997) spectrum with a PL (left) and LP (right). The residuals refer to the curve with the median values of the parameters and the de-absorbed points take into account the uncertainty propagation of EBL dust fractions and their covariance matrix. In gray, 100 random models were sampled from the posterior distribution.}}
\label{fig:mkn501_fit}
\end{figure}

In Appendix \ref{appendix:spec_fit}, we show the fit to all observed data, taking the median values of the respective parameters, alongside 100 random selections from the posterior. We also show the de-absorbed data with propagated uncertainties due to the covariance matrix of EBL parameters. It is not the scope of this paper to do a detailed discussion on each individual source, but we emphasize a few results, listed as follows.
\begin{itemize}
    \item \textbf{Mkn 501 (HEGRA/1997)}: The fit of both PL and LP models, shown in figure \ref{fig:mkn501_fit}, resulted in qualitatively identical residuals, as the intrinsic spectra have virtually the same shape. The median spectral indices for the PL and LP models are $\Gamma = 2.31^{+0.13}_{-0.13}$ and $a = 2.06^{+0.57}_{-0.58}$, respectively, whereas the curvature parameter is $b=0.18^{+0.43}_{-0.39}$. Therefore, the choice of intrinsic spectrum model between these two options does not seem to introduce a relevant systematic error for Mkn 501, although they result in distinct EBL constraints.
    \item \textbf{Mkn 421 (MAGIC/2004--2005)}: \cite{2007Albert} have obtained a photon index $\Gamma = 2.20\pm 0.08$ and a cut-off energy $E_\text{cut}=(1.44\pm 0.27)\si{\tera\electronvolt}$ using a reescaled version of the EBL model from \cite{2005Primack}, both consistent with our results of $\Gamma = 2.242^{+0.080}_{-0.079}$ and $E_\text{cut}=1.64^{+0.43}_{-0.29}\,\si{\tera\electronvolt}$.
    \item \textbf{PKS 2155--304 (H.E.S.S./2006)}: \cite{2013PhRvD..88j2003A}, considering the EBL model from  \cite{2008A&AFranceschini} and the LP parametrization, found a photon index $a = 3.18\pm 0.03\,\text{(stat)}\pm 0.20\,\text{(syst)}$ and a curvature parameter $b = 0.32\pm 0.02\,\text{(stat)}\pm 0.05\,\text{(syst)}$. Although our photon index closely matches their result, with $a= 3.145^{+0.040}_{-0.040}$, we found a relatively higher curvature, of $b = 0.749^{+0.068}_{-0.068}$.
    \item \textbf{PKS 2005--489 (H.E.S.S./2004--2007)}: \cite{2010A&A...511A..52H} found $\Gamma = 2.69\pm 0.16$, based on the EBL model from \cite{2008A&AFranceschini}, which is compatible to our result of $\Gamma = 2.68^{+0.17}_{-0.15}$.
    \item \textbf{3C 279 (MAGIC/2008)}: \cite{2008Sci...320.1752M} present two ``extreme'' values for the intrinsic photon index, of $\Gamma = 2.9\pm0.9\,\text{(stat)}\pm 0.5\,\text{(syst)}$ (low EBL, \cite{2005Primack}) and $\Gamma = 0.5\pm1.2\,\text{(stat)}\pm0.5\,\text{(syst)}$ (high EBL, \cite{2006ApJStecker}). Our result, of $\Gamma = 0.94^{+0.96}_{-0.66}$, is in between both, although closer to the high EBL case. Overall, the uncertainties are large, do not discarding values compatible to $\Gamma \gtrsim 1.5$. Notably, D11 presents a much higher photon index for 3C 279, of $\Gamma = 3.78^{+0.10}_{-0.08}\pm0.88$. In any case, a ``flattening'' of the spectrum in the higher energy bins is identified, as we can see in the bottom-right panel of Figure \ref{fig:spec_fit_PL_6} of Appendix \ref{appendix:spec_fit}.
    \item \textbf{1ES 1011+496 (MAGIC/2007)}: The analysis by \cite{2007ApJ...667L..21A} points to a PL spectrum with photon index $\Gamma = 3.3\pm 0.7$, taking into account the EBL attenuation from \cite{2002A&AKneiske}. Our result suggests a lower value of $\Gamma = 2.87^{+0.63}_{-0.63}$, but still compatible under the uncertainties.
    \item  \textbf{1ES 2344+514 (VERITAS/2007--2008, 2007)}: \cite{2011ApJ...738..169A} estimated intrinsic spectral indices of $\Gamma \approx 2.5$ and $\Gamma \approx 2.1$ for the low and high flux states, respectively, using the EBL model from \cite{2008A&AFranceschini}. Our results are very similar, as we have obtained $\Gamma = 2.46^{+0.10}_{-0.10}$ and $\Gamma = 2.11^{+0.24}_{-0.22}$.
    \item \textbf{1ES 0347--121 (H.E.S.S./2006)}: \cite{HESS:2007yxj} using the EBL model from \cite{2005Primack} and the upper limit curve from \cite{2006Natur.440.1018A}, have obtained $\Gamma = 2.10\pm0.21$ and $\Gamma = 1.69\pm0.22$, respectively. Our result more closely matches the second one, of $\Gamma=1.71^{+0.36}_{-0.32}$.
    \item \textbf{IC 310 (MAGIC/2009--2010)}: \cite{2014A&A...563A..91A} provides $\Gamma =1.81\pm0.13\,\text{(stat)}\pm0.20\,\text{(syst)}$ and $\Gamma = 1.85\pm0.11\,\text{(stat)}\pm0.20\,\text{(syst)}$ for the low and high states, respectively, considering the EBL attenuation from D11. Our result is compatible to both, with the corresponding median values of $\Gamma = 1.80^{+0.17}_{-0.16}$ and $\Gamma=1.81^{+0.11}_{-0.11}$.
\end{itemize}

\section{Conclusions}\label{sec:conclusion}

The EBL has fundamental importance to extragalactic TeV astronomy, as only when attenuation effects in the propagation of gamma rays are taken into account that we can identify the intrinsic spectral properties of the emitting sources. Conversely, with adequate modeling, it is possible to probe the energy distribution of the EBL using gamma-ray observations. We have shown that the Bayesian point of view, combined with MCMC methods, allows us to simultaneously infer EBL and intrinsic spectral parameters. This has the advantage of easily marginalizing over regions of the parameter space, incorporating the uncertainties of the EBL modeling into the source parameters and vice-versa. We have also presented an analytical way of marginalizing over the flux normalization parameters, which can be useful when one wishes to decrease the sample volume in parameter space.

The procedure was applied to a set of synthetic sources observed with the ideal CTA telescope configuration and detected with high significance (TS>25), revealing that the combined fit of a fairly small sample (12 sources) leads to typical resolutions of 4\% for the PAH grain fraction. Such constraints can be further improved when observations are taken with longer exposures ($>$5h). By specifically adopting F10's model with free dust fractions, we could approximately reproduce D11's EBL intensity in the mid-IR. Moreover, as long as the observational time per source is moderate ($<$5h), the intrinsic spectral parameters of the sources can be accurately inferred. Still using synthetic samples, systematic effects due to the incorrect EBL modeling could be identified when sources are observed for long periods ($>$5h), indicating that well-measured spectra can likely help us in the future to distinguish EBL models, particularly in the far-IR range where current models tend to show larger discrepancies among their emissivities. It is expected that, when CTA starts its full operation, the Extragalactic Survey and Active Galactic Nuclei Key Science Projects \cite{CTAConsortium:2017dvg} will greatly contribute to generating the data which can be used in the refinement of EBL constraints. 

Additionally, we considered a sample of 65 spectra from 36 distinct AGNs measured by current IACTs and we have sampled the posterior distribution of EBL and intrinsic parameters using the HMC method. Compared to the previous work of \cite{2019pimentel}, we achieve notably smaller uncertainties for the PAH fraction, as the mid-IR EBL range can be well constrained by the whole data sample. The far-IR region, however, can only be currently constrained by Mkn 501's flare observation (HEGRA/1997), due to a ${\sim}\SI{20}{\tera\electronvolt}$ flux measurement and the relatively small redshift of the source. This means the corresponding EBL result is dependent on the intrinsic model chosen for this spectrum, as the inclusion of an intrinsic curvature with a log-parabolic shape resulted in broader SG and LG distributions. Such choice, nevertheless, leaves the intrinsic parameters of all other sources essentially unaltered, giving us confidence in the robustness of the estimation of the intrinsic spectral parameters.  

The methodology presented here can be naturally extended to other parametric EBL models, or incorporate a more detailed description of the sources' emission, like those based on the SSC paradigm. These extensions are easily accommodated by the Bayesian framework and the versatility of MCMC tools.

\begin{table}[tbp]
    \centering
    \begin{tabular}{c|c|c||c|c|c}
    \hline\hline
        Name & Survey/Year -- Label & Ref. & $N_0$(*)$\times 10^{-6}$ & $\Gamma$ & $E_{\text{cut}}$ (**)\\
        \hline
        \rule{0pt}{3ex} Markarian 421 & MAGIC/2004--2005 & \text{\cite{2007ApJ...663..125A}} & $0.417^{+0.062}_{-0.053}$ & $2.242^{+0.080}_{-0.079}$ & $1.64^{+0.43}_{-0.29}$\\[5pt]
        \rule{0pt}{3ex} & VERITAS/2008 -- High A & \text{\cite{2011ApJ...738...25A}} & $2.53^{+0.39}_{-0.32}$ & $1.81^{+0.12}_{-0.12}$ & $2.06^{+0.59}_{-0.40}$\\[5pt]
        \rule{0pt}{3ex} & VERITAS/2008 -- Low & \text{\cite{2011ApJ...738...25A}} & $1.006^{+0.050}_{-0.047}$ & $2.097^{+0.038}_{-0.038}$ & $3.27^{+0.45}_{-0.36}$\\[5pt]
        \rule{0pt}{3ex} & VERITAS/2008 -- Very Low & \text{\cite{2011ApJ...738...25A}} & $0.62^{+0.10}_{-0.09}$ & $2.11^{+0.11}_{-0.11}$ & $1.76^{+0.49}_{-0.33}$\\[5pt]
        \hline
        \hline
        Name & Survey/Year -- Label & Ref. & $N_0$(*)$\times 10^{-6}$ & $a$ & $b$ \\
        \hline
        \rule{0pt}{3ex} Markarian 421 & VERITAS/2008 -- High B & \text{\cite{2011ApJ...738...25A}} & $2.23^{+0.12}_{-0.13}$ & $2.03^{+0.16}_{-0.17}$ & $0.62^{+0.27}_{-0.26}$\\[5pt]
        \rule{0pt}{3ex} & VERITAS/2008 -- Middle & \text{\cite{2011ApJ...738...25A}} & $1.107^{+0.019}_{-0.019}$ & $2.359^{+0.022}_{-0.022}$ & $0.315^{+0.049}_{-0.051}$\\[5pt]
        \rule{0pt}{3ex} & VERITAS/2008 -- Very High & \text{\cite{2011ApJ...738...25A}} & $3.46^{+0.14}_{-0.14}$ & $1.99^{+0.11}_{-0.12}$ & $0.83^{+0.21}_{-0.20}$\\[5pt]
        \hline
        \rule{0pt}{3ex} Markarian 501 & HEGRA/1997 & \text{\cite{2001A&A...366...62A}} & $1.58^{+0.91}_{-0.59}$ & $2.06^{+0.57}_{-0.58}$ & $0.18^{+0.43}_{-0.39}$ \\[5pt]
        \hline
        \rule{0pt}{3ex} PKS 2155--304 & HESS/2006 & \text{\cite{2013PhRvD..88j2003A}} & $1.256^{+0.021}_{-0.021}$ & $3.145^{+0.040}_{-0.040}$ & $0.749^{+0.068}_{-0.068}$\\[5pt]
    \end{tabular}
    \caption{References of the IACT spectra utilized for PLC and LP sources, alongside the median values and uncertainties (68\% interval) of their respective parameters; (*) in \si{\tera\electronvolt^{-1}\meter^{-2}\second^{-1}} and (**) in \si{\tera\electronvolt}.}
    \label{tab:tevcat_results_curvature}
\end{table}

\begin{table}[tbp]
    \centering
    \begin{tabular}{c|c|c||c|c}
    \hline\hline
        Name & Survey/Year -- Label & Ref. & $N_0$(*)$\times 10^{-8}$ & $\Gamma$\\
        \hline
        \rule{0pt}{3ex} 1ES 0229+200 & HESS/2005--2006 & \cite{HESS:2007xak} & $3.06^{+0.58}_{-0.59}$ & $1.35^{+0.31}_{-0.31}$ \\
        \rule{0pt}{3ex}  &  VERITAS/2010--2012 & \cite{Aliu:2013pya} & $2.57^{+0.33}_{-0.34}$ & $1.59^{+0.17}_{-0.15}$ \\[5pt]
        \hline
        \rule{0pt}{3ex} 1ES 0347--121 & HESS/2006 & \cite{HESS:2007yxj} & $3.81^{+0.71}_{-0.72}$ & $1.71^{+0.25}_{-0.22}$ \\[5pt]
        \hline
        \rule{0pt}{3ex} 1ES 0414+009 & HESS/2005--2009 & \text{\cite{2012A&A...538A.103H}}  & $2.1^{+1.0}_{-0.8}$ & $1.71^{+0.36}_{-0.32}$ \\[5pt]
        \rule{0pt}{3ex} & VERITAS/2008--2011 & \text{\cite{2012ApJ...755..118A}}  & $0.83^{+0.61}_{-0.40}$ & $1.32^{+0.56}_{-0.55}$ \\[5pt]
        \hline
        \rule{0pt}{3ex} 1ES 0806+524 & MAGIC/2011 -- Low & \text{\cite{2015MNRAS.451..739A}}  & $2.8^{+1.9}_{-1.4}$ & $1.71^{+0.60}_{-0.66}$\\[5pt]
        \rule{0pt}{3ex} & MAGIC/2011 -- High & \text{\cite{2015MNRAS.451..739A}} &  $7.2^{+2.8}_{-2.4}$ & $2.05^{+0.31}_{-0.29}$ \\[5pt]
        \rule{0pt}{3ex} & VERITAS/2006--2008 & \text{\cite{2009ApJ...690L.126A}} & $1.9^{+1.9}_{-1.1}$ & $1.8^{+1.0}_{-1.0}$ \\[5pt]
        \hline
        \rule{0pt}{3ex} 1ES 1011+496 & 
        MAGIC/2007 & \text{\cite{2007ApJ...667L..21A}}  & $2.6^{+3.8}_{-1.7}$ & $2.87^{+0.63}_{-0.63}$ \\[5pt]
        \hline
        \rule{0pt}{3ex} 1ES 1101--232 & HESS/2004--2005 & \text{\cite{2006Natur.440.1018A}} & $3.44^{+0.62}_{-0.63}$ & $1.63^{+0.20}_{-0.19}$ \\[5pt]
        \hline
        \rule{0pt}{3ex} 1ES 1215+303 & MAGIC/2011 & \text{\cite{2015MNRAS.451..739A}} & $1.92^{+0.56}_{-0.52}$ & $2.43^{+0.19}_{-0.19}$ \\[5pt]
        \rule{0pt}{3ex} & VERITAS/2011 & \text{\cite{2013ApJ...779...92A}} & $1.17^{+0.40}_{-0.35}$ & $2.68^{+0.40}_{-0.38}$ \\[5pt]
        \hline
        \rule{0pt}{3ex} 1ES 1218+304 & VERITAS/2008--2009 & \text{\cite{2010ApJ...709L.163A}} & $10.0^{+1.2}_{-1.1}$ & $1.91^{+0.10}_{-0.10}$ \\[5pt]
        \rule{0pt}{3ex} & VERITAS/2007 & \text{\cite{2009ApJ...695.1370A}} & $7.1^{+1.9}_{-1.8}$ & $1.68^{+0.37}_{-0.35}$ \\[5pt]
        \rule{0pt}{3ex} & MAGIC/2005 & \text{\cite{2006ApJ...642L.119A}} & $8.7^{+8.5}_{-4.9}$ & $1.88^{+0.59}_{-0.68}$ \\[5pt]
        \hline
        \rule{0pt}{3ex} 1ES 1312--423 & HESS/2004--2010 & \text{\cite{2013MNRAS.434.1889H}} & $0.67^{+0.19}_{-0.19}$ & $1.95^{+0.39}_{-0.36}$ \\[5pt]
        \hline
        \rule{0pt}{3ex} 1ES 1727+502 & VERITAS/2013 & \text{\cite{2015ApJ...808..110A}} & $4.74^{+0.93}_{-0.95}$ & $1.76^{+0.30}_{-0.29}$ \\[5pt]
        \hline
        \rule{0pt}{3ex} 1ES 1959+650 & VERITAS/2007--2011 & \text{\cite{2013ApJ...775....3A}} & $10.11^{+0.87}_{-0.89}$ & $2.14^{+0.097}_{-0.099}$ \\[5pt]
        \rule{0pt}{3ex} & MAGIC/2006 & \text{\cite{2008ApJ...679.1029T}} & $4.28^{+0.57}_{-0.57}$ & $2.21^{+0.18}_{-0.17}$ \\[5pt]
        \hline
        \rule{0pt}{3ex} 1ES 2344+514 & VERITAS/2007--2008 & \text{\cite{2011ApJ...738..169A}} & $5.26^{+0.36}_{-0.36}$ & $2.46^{+0.10}_{-0.10}$ \\[5pt]
        \rule{0pt}{3ex} & VERITAS/2007 -- High & \text{\cite{2011ApJ...738..169A}} & $26.5^{+3.1}_{-3.1}$ & $2.11^{+0.24}_{-0.22}$ \\[5pt]
        \rule{0pt}{3ex} & MAGIC/2005--2006 & \text{\cite{2007ApJ...662..892A}} & $2.66^{+0.42}_{-0.43}$ & $2.69^{+0.13}_{-0.13}$ \\[5pt]
        \hline
        \rule{0pt}{3ex} 1RXS J101015.9 & HESS/2006--2010 & \text{\cite{2012A&A...542A..94H}} & $0.94^{+0.36}_{-0.36}$ & $1.96^{+0.53}_{-0.45}$ \\[5pt]
        \hline
        \rule{0pt}{3ex} 3C 279 & MAGIC/2008 & \text{\cite{2008Sci...320.1752M}}  & $370^{+610}_{-290}$ & $0.94^{+0.96}_{-0.66}$ \\[5pt]
        \hline
    \end{tabular}
    \caption{References of the IACT spectra utilized for PL sources, alongside the median values and uncertainties (68\% interval) of their respective parameters; (*) in \si{\tera\electronvolt^{-1}\meter^{-2}\second^{-1}}.}
    \label{tab:tevcat_results}
\end{table}

\begin{table}[tbp]
    \centering
    \begin{tabular}{c|c|c||c|c}
    \hline\hline
        Name & Survey/Year -- Label & Ref. & $N_0$(*)$\times 10^{-8}$ & $\Gamma$\\
        \hline
        \rule{0pt}{3ex} 3C 66A & VERITAS/2008 -- Low & \text{\cite{2011ApJ...726...43A}} & $14^{+16}_{-8}$ & $1.93^{+0.57}_{-0.57}$ \\[5pt]
        \rule{0pt}{3ex} & VERITAS/2008 -- High & \text{\cite{2011ApJ...726...43A}} & $28^{+36}_{-16}$ & $1.87^{+0.63}_{-0.65}$ \\[5pt]
        \hline
        \rule{0pt}{3ex} 4C +21.35 & MAGIC/2010 & \text{\cite{2011ApJ...730L...8A}} & $46^{+31}_{-20}$ & $2.39^{+0.28}_{-0.27}$ \\[5pt]
        \hline
        \rule{0pt}{3ex} AP Librae & HESS/2010--2011 & \text{\cite{2015A&A...573A..31H}} & $0.75^{+0.15}_{-0.15}$ & $2.35^{+0.19}_{-0.17}$ \\[5pt]
        \hline
        \rule{0pt}{3ex} BL Lacertae & VERITAS/2011 & \text{\cite{2013ApJ...762...92A}} & $12.9^{+6.8}_{-5.0}$ & $3.30^{+0.38}_{-0.36}$ \\[5pt]
        \hline
        \rule{0pt}{3ex} Centaurus A & HESS/2004--2008 & \text{\cite{2009ApJ...695L..40A}} &  $0.241^{+0.056}_{-0.058}$ & $2.77^{+0.48}_{-0.41}$ \\[5pt]
        \hline
        \rule{0pt}{3ex} H 1426+428 & HEGRA/1999--2000 & \text{\cite{2003A&A...403..523A}} & $13.4^{+8.6}_{-6.2}$ & $1.9^{+1.2}_{-0.8}$ \\[5pt]
        \rule{0pt}{3ex} & HEGRA/2002 & \text{\cite{2003A&A...403..523A}} & $2.0^{+1.5}_{-0.9}$ & $0.93^{+0.55}_{-0.51}$ \\[5pt]
        \hline
        \rule{0pt}{3ex} H 2356--309 & HESS/2004--2007 & \text{\cite{2010A&A...516A..56H}} & $2.02^{+0.28}_{-0.28}$ & $1.90^{+0.16}_{-0.15}$ \\[5pt]
        \hline
        \rule{0pt}{3ex} IC 310 & MAGIC/2012 & \text{\cite{2014Sci...346.1080A}} & $21.6^{+1.7}_{-1.6}$ & $1.776^{+0.067}_{-0.064}$ \\[5pt]
        \rule{0pt}{3ex} & MAGIC/2009--2010 -- High & \text{\cite{2014A&A...563A..91A}} & $4.97^{+0.53}_{-0.52}$ & $1.81^{+0.11}_{-0.11}$ \\[5pt]
        \rule{0pt}{3ex} & MAGIC/2009--2010 -- Low & \text{\cite{2014A&A...563A..91A}} & $0.67^{+0.12}_{-0.13}$ & $1.80^{+0.17}_{-0.16}$\\[5pt]
        \hline
        \rule{0pt}{3ex} M87 & HESS/2005 & \text{\cite{2006Sci...314.1424A}} & $1.22^{+0.16}_{-0.16}$ & $2.16^{+0.15}_{-0.14}$ \\[5pt]
        \rule{0pt}{3ex} & HESS/2004 & \text{\cite{2006Sci...314.1424A}}  & $0.28^{+0.10}_{-0.10}$ & $2.43^{+0.46}_{-0.41}$ \\[5pt]
        \rule{0pt}{3ex} & MAGIC/2005--2007 & \text{\cite{2012A&A...544A..96A}} & $0.56^{+0.12}_{-0.11}$ & $2.15^{+0.21}_{-0.22}$ \\[5pt]
        \rule{0pt}{3ex} & MAGIC/2008 & \text{\cite{2008ApJ...685L..23A}} & $2.88^{+0.42}_{-0.43}$ & $2.28^{+0.12}_{-0.11}$\\[5pt]
        \rule{0pt}{3ex} & VERITAS/2007 & \text{\cite{2008ApJ...679..397A}}  & $0.74^{+0.13}_{-0.14}$ & $2.27^{+0.19}_{-0.17}$ \\[5pt]
        \hline
        \rule{0pt}{3ex} Markarian 180 & MAGIC/2006 & \text{\cite{2006ApJ...648L.105A}} & $1.51^{+0.83}_{-0.74}$ & $2.52^{+0.65}_{-0.88}$ \\[5pt]
        \hline
        \rule{0pt}{3ex} Markarian 421 & MAGIC/2006 & \text{\cite{2009ApJ...703..169A}}  & $33.9^{+2.3}_{-2.4}$ & $2.068^{+0.088}_{-0.084}$ \\[5pt]
        \rule{0pt}{3ex} & VERITAS/2008 -- High C & \text{\cite{2011ApJ...738...25A}}  & $276^{+17}_{-17}$ & $2.422^{+0.090}_{-0.085}$\\[5pt]
    \end{tabular}
    \caption{Same as table~\ref{tab:tevcat_results}; (*) in \si{\tera\electronvolt^{-1}\meter^{-2}\second^{-1}}.}
    \label{tab:tevcat_results2}
\end{table}

\begin{table}[tbp]
    \centering
    \begin{tabular}{c|c|c||c|c}
    \hline\hline
        Name & Survey/Year -- Label & Ref. & $N_0$(*)$\times 10^{-8}$ & $\Gamma$\\
        \hline
        \rule{0pt}{3ex} Markarian 501 & VERITAS/2009 & \text{\cite{2011ApJ...729....2A}}  & $8.1^{+1.1}_{-1.2}$ & $2.49^{+0.16}_{-0.15}$\\[5pt]
        \hline
        \rule{0pt}{3ex} NGC 1275 & MAGIC/2009--2014 & \text{\cite{2016A&A...589A..33A}}  & $0.078^{+0.024}_{-0.021}$ & $3.45^{+0.17}_{-0.16}$ \\[5pt]
        \hline
        \rule{0pt}{3ex} PKS 0447--439 & HESS/2009 & \text{\cite{2013A&A...552A.118H}} & $24.5^{+7.9}_{-7.4}$ & $1.20^{+0.37}_{-0.33}$ \\[5pt]
        \hline
        \rule{0pt}{3ex} PKS 1441+25 & MAGIC/2015 & \text{\cite{2015ApJ...815L..23A}} & $5.3^{+2.5}_{-1.7}$ & $2.80^{+0.16}_{-0.16}$ \\[5pt]
        \hline
        \rule{0pt}{3ex} PKS 1510--089 & HESS/2009 & \text{\cite{2013A&A...554A.107H}} & $2.2^{+9.3}_{-1.8}$ & $2.4^{+1.1}_{-1.2}$\\[5pt]
        \hline
        \rule{0pt}{3ex} PKS 2005--489 & HESS/2004--2007 & \text{\cite{2010A&A...511A..52H}}  & $1.48^{+0.13}_{-0.14}$ & $2.68^{+0.17}_{-0.15}$ \\[5pt]
        \hline
        \rule{0pt}{3ex} PKS 2155--304 & HESS/2005--2007 & \text{\cite{2010A&A...520A..83H}}  & $5.84^{+0.37}_{-0.37}$ & $2.831^{+0.056}_{-0.054}$ \\[5pt]
        \rule{0pt}{3ex} & MAGIC/2006 & \text{\cite{2012A&A...544A..75A}}& $59.9^{+6.3}_{-6.6}$ & $2.42^{+0.12}_{-0.11}$ \\[5pt]
        \hline
        \rule{0pt}{3ex} RBS 0413 & VERITAS/2009 & \text{\cite{2012ApJ...750...94A}} & $3.1^{+1.9}_{-1.4}$ & $1.46^{+0.71}_{-0.69}$ \\[5pt]
        \hline
        \rule{0pt}{3ex} RGB J0152+017 & HESS/2007 & \text{\cite{2008A&A...481L.103A}} & $1.31^{+0.34}_{-0.37}$ & $2.36^{+0.38}_{-0.32}$ \\[5pt]
        \hline
        \rule{0pt}{3ex} RGB J0710+591 & VERITAS/2008--2009 & \text{\cite{2010ApJ...715L..49A}} & $3.31^{+0.65}_{-0.67}$ & $1.80^{+0.28}_{-0.27}$ \\[5pt]
        \hline
        \rule{0pt}{3ex} RX J0648.7+1516 & VERITAS/2010 & \text{\cite{2011ApJ...742..127A}} & $1.3^{+1.7}_{-0.8}$ & $2.79^{+0.76}_{-0.74}$ \\[5pt]
    \end{tabular}
    \caption{Same as table~\ref{tab:tevcat_results}; (*) in \si{\tera\electronvolt^{-1}\meter^{-2}\second^{-1}}.}
    \label{tab:tevcat_results3}
\end{table}

\acknowledgments

The authors acknowledge the helpful comments and suggestions from Amy Furniss and Jonathan Biteau, as well as the Consortium Speakers and Publications Office of CTA for organizing the review process. M.G., E.M.S. and L.A.S. acknowledge Funda\c{c}\~ao de Amparo \`a Pesquisa do Estado de S\~ao Paulo (FAPESP) for Projeto Tem\'atico grant numbers 2015/15897-1 and  2021/01089-1. M.G. was also supported by FAPESP grant number 2021/01473-6 and Conselho Nacional de Desenvolvimento Cient\'ifico e Tecnol\'ogico (CNPq) grant number 130534/2021-8. L.A.S. was also supported by FAPESP grant number 	21/00030-3. E.M.S. is grateful to CNPq for grant number 310125/2019-7. This research was developed with HPC resources from Superintendência de Tecnologia da Informação of University of S\~ao Paulo. This research made use of ctools, a community-developed analysis package for Imaging Air Cherenkov Telescope data. ctools is based on GammaLib, a community-developed toolbox for the scientific analysis of astronomical gamma-ray data.

\appendix
\section{Marginalization of the likelihood over the flux normalisation}\label{appendix:marginal}
Our goal is to marginalise the posterior distribution over the flux normalisation of each source participating in the likelihood, by computing 
\begin{equation}
    p(\vb{\Omega}_r|D,I) = \int^\infty_{-\infty}\cdots\int^\infty_{-\infty}\left(\prod^N_{j=1}\dd{N^{(j)}_0}\right)\frac{p(D|\vb{\Omega},I)p(\vb{\Omega}|I)}{p(D|I)},
\end{equation}
where $\vb{\Omega}_r$ is the set of parameters after integrating out the normalization variables of $N$ sources (each labelled as $j$). We also use an improper prior for which $p(\Omega|I)=0$ if any $N^{(j)}_0 < 0$. As the prior and evidence are constants during the integration, our task is to marginalize the likelihood 
\begin{equation}
    p(D|\vb{\Omega}_r,I) = \int^\infty_{0}\cdots\int^\infty_0\left(\prod^N_{j=1}\dd{N^{(j)}_0}\right)\frac{1}{Z}\exp(-\frac{1}{2}\chi^2),
\end{equation}
where $Z$ is the probability normalisation and 
\begin{equation}
    \chi^2 = \sum^N_{j=1}\sum^{n_j}_{i=1}\left[\frac{\Phi^{(j)}_{\rm obs}\left(E^{(j)}_i\right)-\Phi^{(j)}_{\rm mod}\left(E^{(j)}_i;\vb{\Omega}\right)}{\sigma\left(E^{(j)}_i\right)}\right]^2.
\end{equation}
In general, the spectral model (for PL, LP or PLC) can be written as
\begin{align}
    \Phi^{(j)}_{\text{mod}}(E;\vb{\Omega})=N^{(j)}_0\tilde{\phi}_{\text{mod}}(E;\vb{\Omega}_r),
\end{align}
where $\tilde{\phi}_{\text{mod}}(E;\vb{\Omega}_r)$ is the energy dependent part. Due to independence of the normalization variables, we arrive at a product of integrals
\begin{equation}
    p(D|\vb{\Omega}_r,I) =\frac{1}{Z}\prod^N_{j=1}\left\{\int^\infty_0\dd{N^{(j)}_0\exp[-\frac{1}{2}\sum^{n_j}_{i=1}\left(\frac{\Phi^{(j)}_{\rm obs}\left(E^{(j)}_i\right)-N^{(j)}_0\tilde{\phi}^{(j)}_{\rm mod}\left(E^{(j)}_i;\vb{\Omega}_r\right)}{\sigma\left(E^{(j)}_i\right)}\right)^2]} \right\}.
\end{equation}
For each integral, by completing the square, a Gaussian integral can be performed. Simplifying the notation, we may realize that
\begin{equation}
    \sum_i\left(\frac{\Phi(E_i)-N_0\tilde{\phi}(E_i)}{\sigma(E_i)} \right)^2 = \sum_i\frac{N^2_0\tilde{\phi}^2(E_i)-2N_0\Phi(E_i)\tilde{\phi}(E_i)+\Phi^2(E_i)}{\sigma^2(E_i)}
\end{equation}
resulting in
\begin{align}
    AN^2_0-2N_0B+C,
\end{align}
where we define
\begin{align}
    A &\equiv \left(\sum_i\frac{\tilde{\phi}^2(E_i)}{\sigma^2(E_i)} \right)\\
    B &\equiv \left(\sum_i\frac{\Phi(E_i)\tilde{\phi}(E_i)}{\sigma^2(E_i)} \right)\\
    C&\equiv \left(\sum_i\frac{\Phi^2(E_i)}{\sigma^2(E_i)} \right),
\end{align}
which assume different values for each source $j$. By completing the square,
\begin{equation}
    AN^2_0-2N_0B+C = A\left(N^2_0-2N_0\frac{B}{A}+\frac{C}{A}\right) = A\left(N_0-\frac{B}{A}\right)^2 +C-\frac{B^2}{A},
\end{equation}
so the marginal likelihood becomes
\begin{equation*}
    p(D|\vb{\Omega}_r,I)=\frac{1}{Z}\prod^N_{j=1}\left\{\int^\infty_0\dd{N^{(j)}_0}\exp[-\frac{A_j}{2}\left(N^{(j)}_0-\frac{B_j}{A_j}\right)^2]\exp(-\frac{1}{2}\left(C_j-\frac{B^2_j}{A_j}\right)) \right\}
\end{equation*}
and is simplified to
\begin{equation*}\label{eq:marginal_like_step}
    p(D|\vb{\Omega}_r,I)=\frac{1}{Z}\exp[-\frac{1}{2}\sum^N_{j=1}\left(C_j-\frac{B^2_j}{A_j}\right)]\prod^N_{j=1}\left\{\int^\infty_0\dd{N^{(j)}_0}\exp[-\frac{A_j}{2}\left(N^{(j)}_0-\frac{B_j}{A_j}\right)^2] \right\}.
\end{equation*}
Changing variables and performing the analytical Gaussian integral, we arrive at
\begin{equation}
    p(D|\vb{\Omega}_r,I)=\frac{1}{Z}\exp[-\frac{1}{2}\sum^N_{j=1}\left(C_j-\frac{B^2_j}{A_j}\right)]\prod^N_{j=1}\sqrt{\frac{\pi}{2A_j}}\left[1+ \erf\left(\frac{B_j}{\sqrt{2A_j}}\right)\right],
\end{equation}
which can be computationally implemented for MCMC purposes. 

\section{Statistical properties of MCMC simulations}\label{appendix:ebl_params}

Table \ref{tab:results_stats_F10} presents the statistical properties of the MCMC simulations using synthetic samples. The acceptance fraction (a.f.) consists on the fraction of accepted proposed steps during the Markov chain, while ESS is the effective sample size after removing the burn-in phase and collecting only steps separated by the integrated autocorrelation time (as presented by \cite{sokal1997monte} and \cite{2013emcee}). To control the Monte Carlo error (the fact that samples presents some degree of correlation), we follow \cite{2015vats} and \cite{gong2016practical} to produce a sufficient large ESS such that the confidence intervals of any parameter is smaller than a fraction $\epsilon$ of the variance of the posterior sample. Fixing the confidence interval in 95\%, table~\ref{tab:results_stats_F10} shows the corresponding value of the precision $\epsilon$, as computed from the formula described by \cite{2013flegal}.

For the HMC simulations, we achieved 11075 samples in the case of Mkn 501 modelled as PL ($\epsilon=4.22\%$); 11539 samples with the LP model ($\epsilon=4.13\%$) and 6195 effective samples for the simulation without Markarian 501 flare data ($\epsilon=5.64\%$).

\begin{table}[H]
    \centering
    \begin{tabular}{c|c|c|c||c|c|c}
    \hline
    \multicolumn{7}{c}{True EBL: F10+P19 } \\
    \hline
    \multicolumn{1}{c}{} & \multicolumn{3}{c||}{Obs. Time $< 5$\,h} & \multicolumn{3}{c}{Obs. Time $\geq 5$\,h}\\\hline
        N.S. & a.f & ESS & $\epsilon$ & a.f. & ESS & $\epsilon$ \\\hline
        2 & 0.654 & \num{3e5} & \num{8.38e-3} & 0.644 & \num{1.8e5} & \num{1.08e-2} \\
        3 & 0.559 & \num{1.8e5} & \num{1.09e-2} & 0.555 & \num{1.2e5} & \num{1.34e-2}\\
        4 &  0.485 & \num{2.25e5} & \num{9.84e-3} & 0.486 & \num{1.125e5} & \num{1.39e-2}\\
        5 &  0.428 & \num{2.7e5} & \num{9.01e-3} & 0.428 & \num{1.35e5} & \num{1.27e-2} \\
        6 &  0.371 & \num{2.1e5} & \num{1.02e-2} & 0.377 & \num{1.099e5}& \num{1.42e-2}\\
        7 & 0.337 & \num{2e5} & \num{1.05e-2} & 0.335 & \num{1.314e5}& \num{1.30e-2}\\
        8 & 0.285 & \num{2e5} & \num{1.05e-2} & 0.290 & \num{1e5} & \num{1.49e-2}\\
        9 & 0.263 & \num{2.4e5} & \num{9.59e-3} & 0.254 & \num{1.2e5} & \num{1.36e-2}\\
        10 & 0.224 & \num{1.995e5} & \num{1.05e-2} & 0.225 & \num{1.08e5} & \num{1.43e-2}\\
        11 & 0.190 & \num{1.71e5} & \num{1.14e-2} & 0.201 & \num{1.098e5}& \num{1.42e-2}\\
        12 & 0.173 & \num{2e5} & \num{1.05e-2} & 0.174 & \num{1.2e5} & \num{1.35e-2}\\
        \hline\hline
        \multicolumn{7}{c}{True EBL: D11} \\
        \hline      
    \multicolumn{1}{c}{} & \multicolumn{3}{|c||}{Obs. Time $< 5$\,h} & \multicolumn{3}{c}{Obs. Time $\geq 5$\,h}\\\hline
        N.S. & a.f & ESS & $\epsilon$ & a.f. & ESS & $\epsilon$ \\\hline
        2 & 0.662 & \num{3.6e5} & \num{7.65e-3} & 0.663 & \num{3.6e5} & \num{7.65e-3}\\
        3 & 0.568 & \num{2.248e5} & \num{9.78e-3} & 0.573 & \num{1.8e5} & \num{1.09e-2}\\
        4 &  0.477 & \num{1.5e5} & \num{1.20e-2} & 0.487 & \num{1.8e5} & \num{1.10e-2}\\
        5 &  0.432 & \num{1.758e5} & \num{1.12e-2} & 0.429 & \num{1.5e5} & \num{1.21e-2}\\
        6 &  0.385 & \num{1.68e5} & \num{1.14e-2} & 0.396 & \num{1.76e5}& \num{1.12e-2}\\
        7 & 0.340 & \num{2e5} & \num{1.05e-2} & 0.336 & \num{1.4e5}& \num{1.26e-2}\\
        8 & 0.288 & \num{1.42e5} & \num{1.25e-2} & 0.283 & \num{1.16e5} & \num{1.38e-2}\\
        9 & 0.254 & \num{1.368e5} & \num{1.27e-2} & 0.252 & \num{1.14e5} & \num{1.39e-2}\\
        10 & 0.227 & \num{1.71e5} & \num{1.14e-2} & 0.221 & \num{1.188e5} & \num{1.36e-2}\\
        11 & 0.192 & \num{1.2e5} & \num{1.36e-2} & 0.197 & \num{1.2e5}& \num{1.36e-2}\\
        12 & 0.169 & \num{1.152e5} & \num{1.38e-2} & 0.174 & \num{1.188e5} & \num{1.36e-2}\\
        \hline
    \end{tabular}
    \caption{MCMC information about the analysis performed with simulated spectra with EBL absorption from \cite{2019pimentel} and D11. For each MCMC run we present the number of sources (N.S.), the acceptance fraction (a.f.), the effective sample size (ESS) and the precision, or Monte Carlo error, $\epsilon$, adopting a confidence interval of $95\%$. }
    \label{tab:results_stats_F10}
\end{table}

\section{Spectral fit of IACT data}\label{appendix:spec_fit}

\begin{figure}[t]
    \centering
    \includegraphics[width=\textwidth]{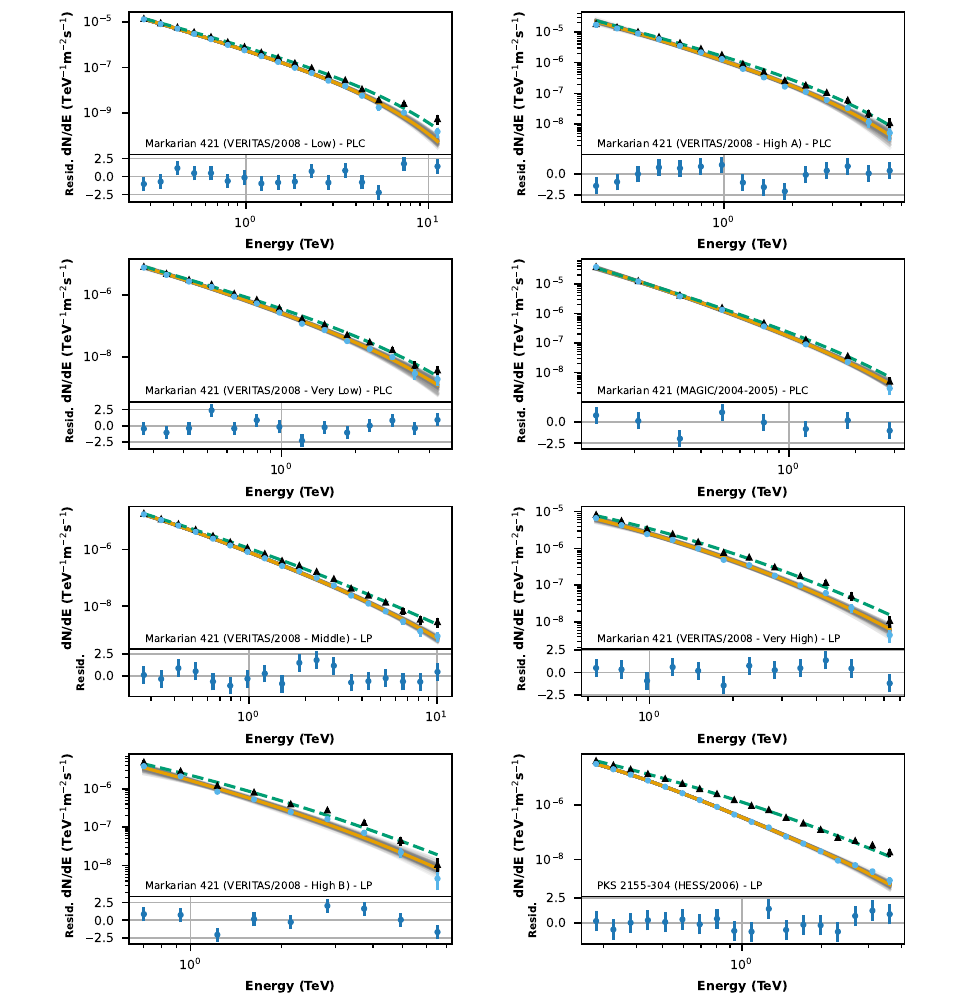}
    \caption{Spectrum fit of sources used in the HMC sampling. Blue circles are the data points and the vermilion solid line is the fit using the median values of the marginalized posterior distributions. The residuals of the fit are shown below the spectrum. The black triangles are the de-absorbed data points with uncertainty propagation taking into account the covariance matrix of dust grains, while the dashed green line is the intrinsic spectrum. We also plot 100 models randomly sampled from the posterior, in gray.}
    \label{fig:spec_fit_PLC_LP}
\end{figure}

\begin{figure}[t]
    \centering
    \includegraphics[width=\textwidth]{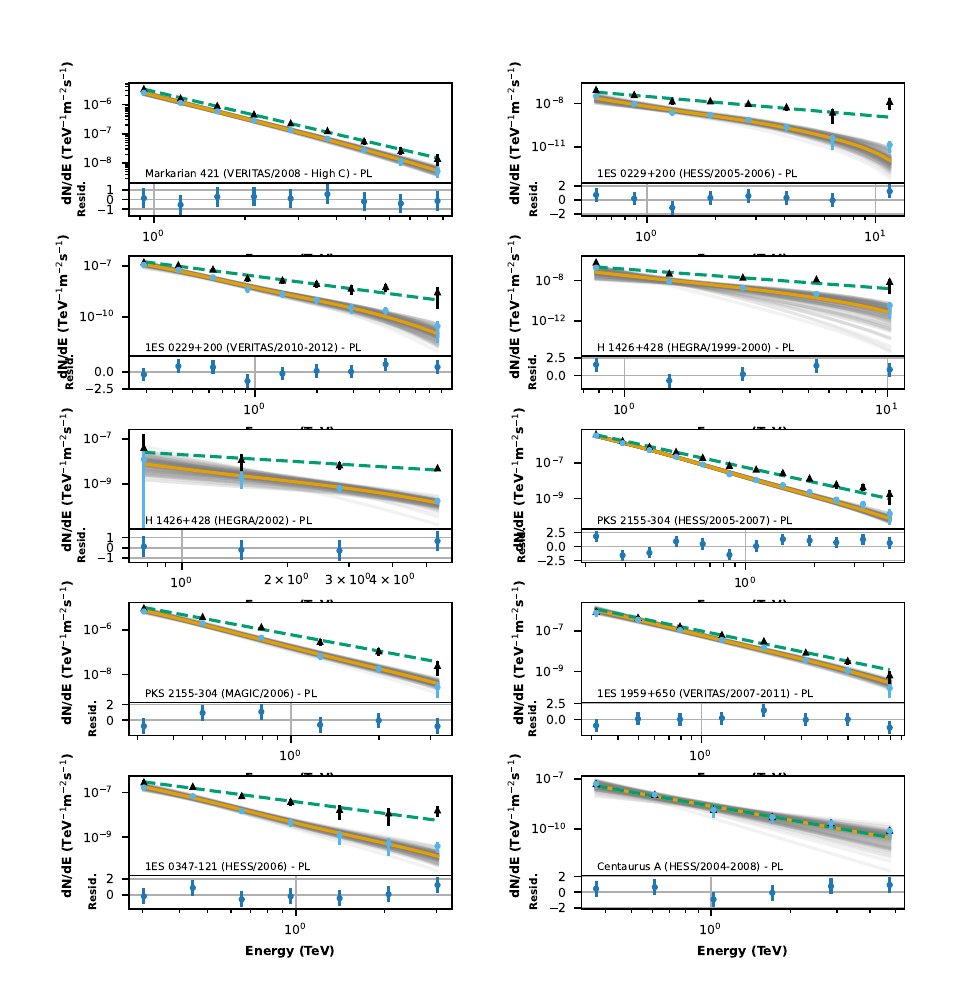}
    \caption{Same as figure~\ref{fig:spec_fit_PLC_LP}.}
    \label{fig:spec_fit_PL_1}
\end{figure}

\begin{figure}[t]
    \centering
    \includegraphics[width=\textwidth]{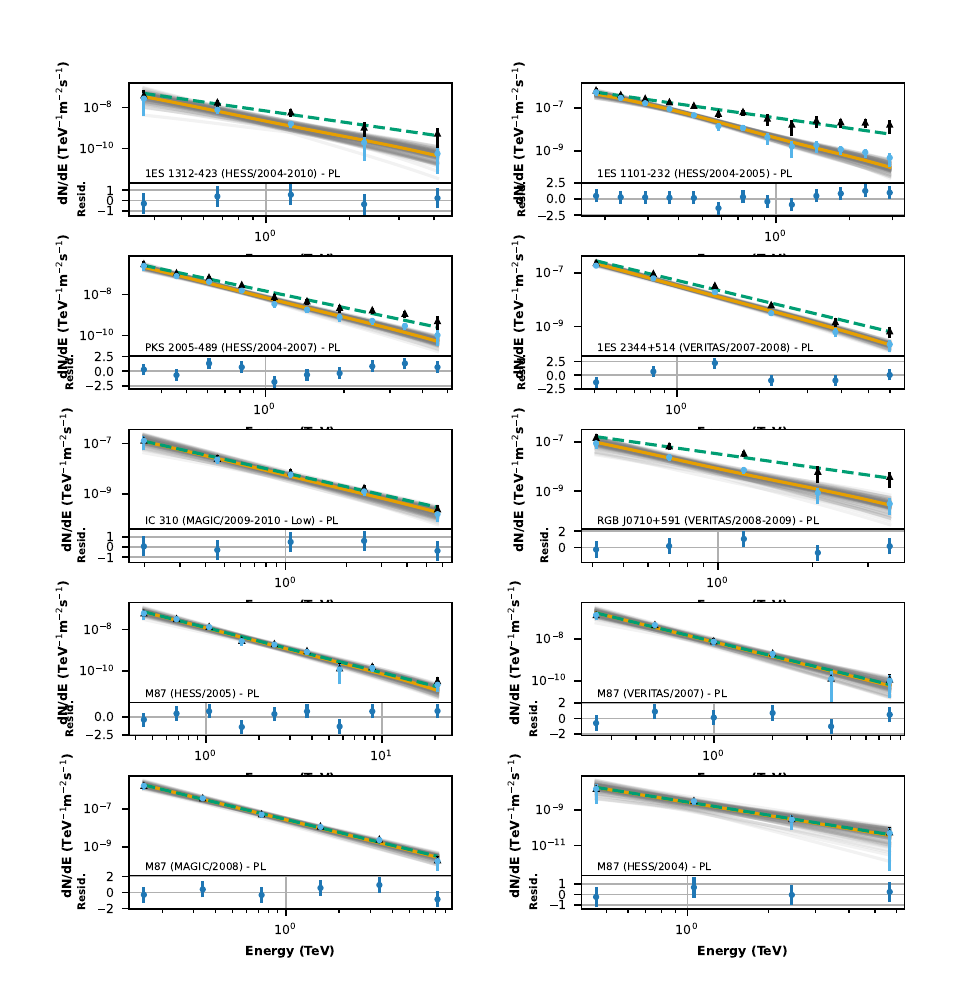}
    \caption{Same as figure~\ref{fig:spec_fit_PLC_LP}.}
    \label{fig:spec_fit_PL_2}
\end{figure}

\begin{figure}[t]
    \centering
    \includegraphics[width=\textwidth]{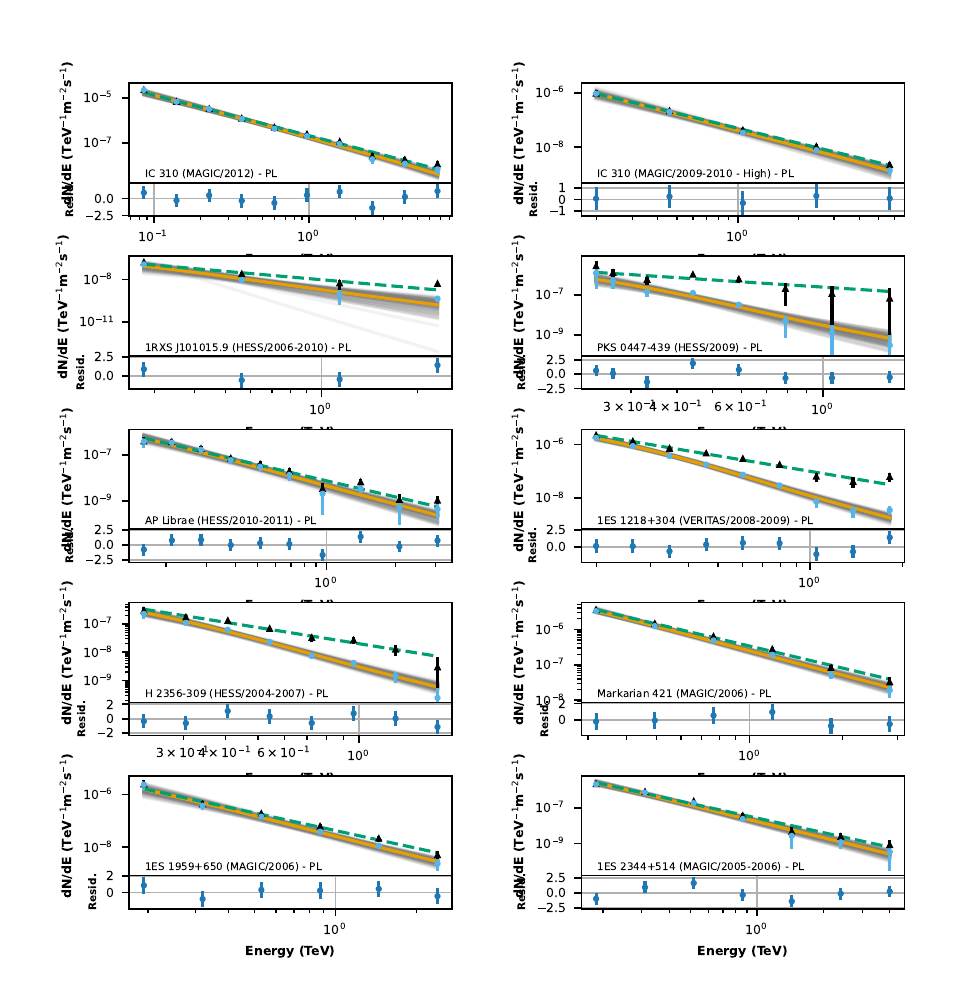}
    \caption{Same as figure~\ref{fig:spec_fit_PLC_LP}.}
    \label{fig:spec_fit_PL_3}
\end{figure}

\begin{figure}[t]
    \centering
    \includegraphics[width=\textwidth]{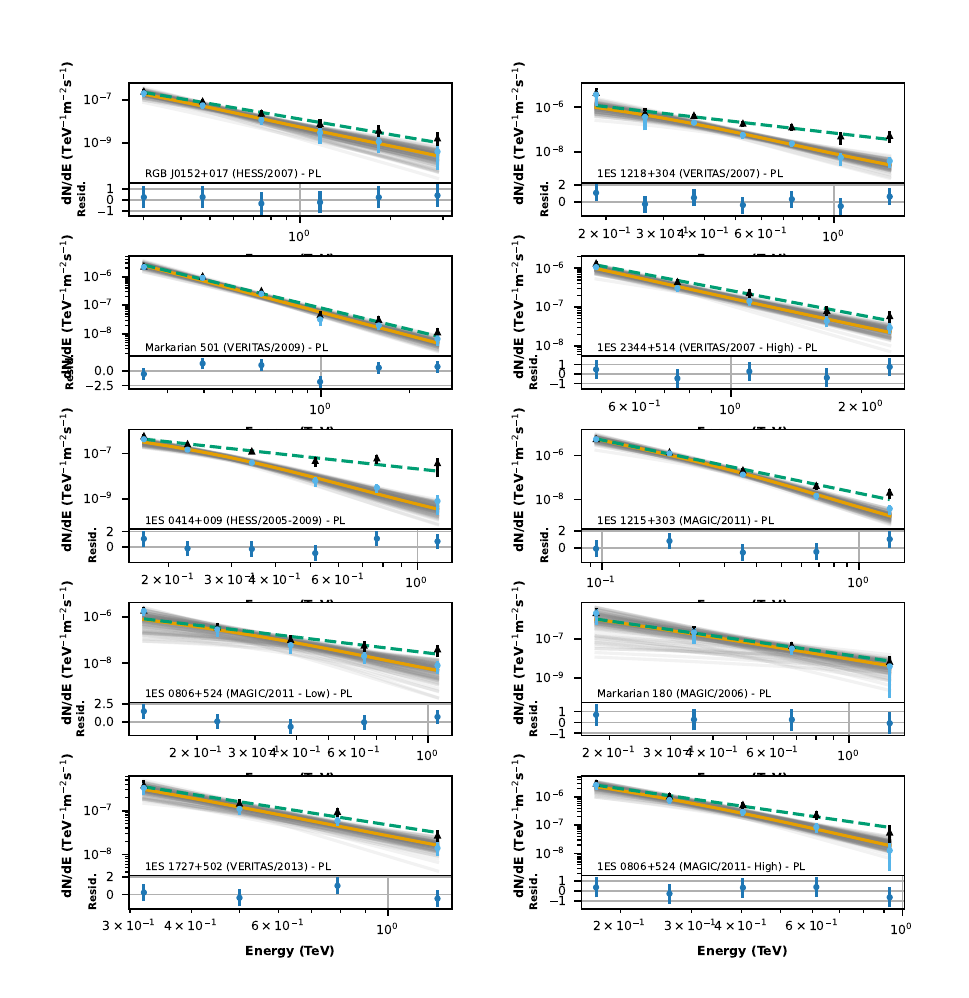}
    \caption{Same as figure~\ref{fig:spec_fit_PLC_LP}.}
    \label{fig:spec_fit_PL_4}
\end{figure}

\begin{figure}[t]
    \centering
    \includegraphics[width=\textwidth]{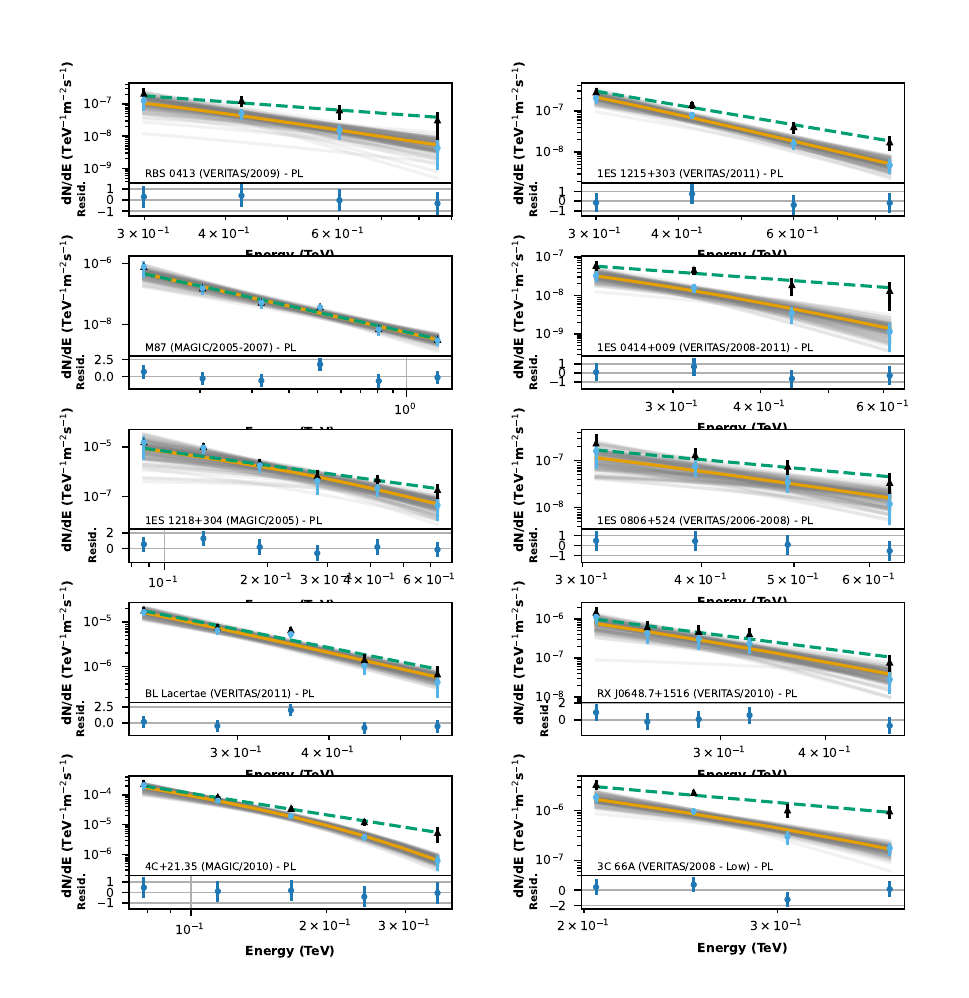}
    \caption{Same as figure~\ref{fig:spec_fit_PLC_LP}.}
    \label{fig:spec_fit_PL_5}
\end{figure}

\begin{figure}[t]
    \centering
    \includegraphics[width=\textwidth]{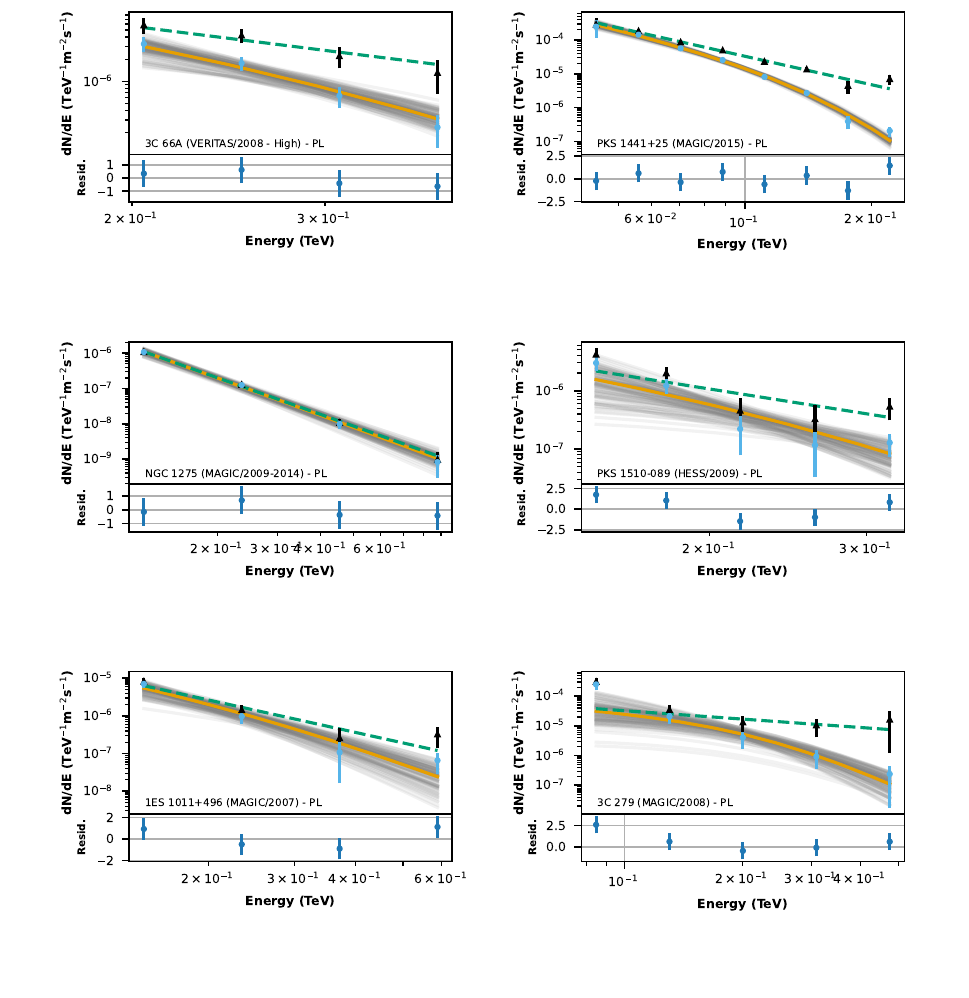}
    \caption{Same as figure~\ref{fig:spec_fit_PLC_LP}.}
    \label{fig:spec_fit_PL_6}
\end{figure}

\bibliographystyle{JHEP}
\bibliography{ref}

\end{document}